\documentclass[12pt]{article}
\usepackage{amsmath,amssymb}
\usepackage{graphicx}

\newcommand{\eqdef}{\stackrel{\text{def}}{=}}
\newcommand{\n}{\nonumber \\}
\newcommand{\bm}{\boldsymbol}
\newcommand{\ignore}[1]{}
\renewcommand{\theequation}{\arabic{section}.\arabic{equation}}

\allowdisplaybreaks[4]

\begin{document}

\baselineskip=20pt

\newfont{\elevenmib}{cmmib10 scaled\magstep1}
\newcommand{\preprint}{
     \begin{flushleft}
       \elevenmib Yukawa\, Institute\, Kyoto\\
     \end{flushleft}\vspace{-1.3cm}
     \begin{flushright}\normalsize  \sf
       DPSU-08-1\\
       YITP-08-1\\
       February 2008
     \end{flushright}}
\newcommand{\Title}[1]{{\baselineskip=26pt
     \begin{center} \Large \bf #1 \\ \ \\ \end{center}}}
\newcommand{\Author}{\begin{center}
     \large \bf Satoru Odake${}^a$ and Ryu Sasaki${}^b$ \end{center}}
\newcommand{\Address}{\begin{center}
       $^a$ Department of Physics, Shinshu University,\\
       Matsumoto 390-8621, Japan\\
       ${}^b$ Yukawa Institute for Theoretical Physics,\\
       Kyoto University, Kyoto 606-8502, Japan
     \end{center}}
\newcommand{\Accepted}[1]{\begin{center}
     {\large \sf #1}\\ \vspace{1mm}{\small \sf Accepted for Publication}
     \end{center}}

\preprint
\thispagestyle{empty}
\bigskip\bigskip\bigskip

\Title{Exactly solvable `discrete' quantum mechanics;
shape invariance, Heisenberg solutions,
annihilation\hspace{0pt}-creation operators and coherent states}
\Author

\Address
\vspace{1cm}

\begin{abstract}
Various examples of exactly solvable `discrete' quantum mechanics
are explored explicitly with emphasis on shape invariance,
Heisenberg operator solutions, annihilation-creation operators,
the dynamical symmetry algebras and coherent states.
The eigenfunctions are the ($q$-)Askey-scheme of hypergeometric
orthogonal polynomials satisfying difference equation versions
of the Schr\"odinger equation. Various reductions (restrictions)
of the symmetry algebra of the Askey-Wilson system are explored
in detail.
\end{abstract}

\section{Introduction}
\label{intro}

General theory of exactly solvable `discrete' quantum mechanics
of one degree of freedom systems is presented with all known examples.
The `discrete' quantum mechanics is a simple extension or deformation
of quantum mechanics in which the momentum operator $p$ appears in
the Hamiltonian in the exponentiated forms $e^{\pm\gamma p}$,
$\gamma\in\mathbb{R}$, in stead of polynomials in ordinary
quantum mechanics. The corresponding Schr\"odinger equations are
difference equations with imaginary shifts, in stead of differential.
The eigenfunctions of the exactly solvable `discrete' quantum mechanics
of one degree of freedom systems consist of the ($q$-)Askey-scheme
of hypergeometric orthogonal polynomials \cite{askey,ismail},
which are deformations of the classical orthogonal polynomials, like
the Hermite, Laguerre, Jacobi polynomials, etc \cite{szego},
constituting the eigenfunctions of exactly solvable ordinary quantum mechanics
\cite{infhul,susyqm}. These eigenpolynomials are orthogonal with respect
to absolutely continuous measure functions, which are just the square
of the ground state wavefunctions; a familiar situation in quantum
mechanics. For another type of orthogonal polynomials with discrete
measures \cite{askey,ismail,koeswart}, see \cite{os12} for a unified
theory. Like most exactly solvable quantum mechanics, every example of
exactly solvable `discrete' quantum mechanics is endowed with dynamical
symmetry, {\em shape invariance\/} \cite{genden}, which allows to
determine the entire  energy spectrum and the corresponding
eigenfunctions when combined with Crum's theorem \cite{crum} or the
factorisation method \cite{infhul,susyqm}.
In other words, shape invariance guarantees exact solvability in the
Schr\"odinger picture \cite{os4,os5,os6}.
As expected, exact solvability in the Heisenberg picture also holds
for all these examples. The explicit forms of Heisenberg operator
solutions give rise to the explicit expressions of annihilation/creation
operators as the positive/negative frequency parts \cite{os7}.
The annihilation/creation operators together with the Hamiltonian constitute
the dynamical symmetry algebra.
In some cases, the algebras are simple and tangible, like the
oscillator algebra and its $q$-deformations \cite{os11}, or
$\mathfrak{su}(1,1)$.

The present paper is to supplement or to complete some results in
previous publications \cite{os4,os5,os6,os7}.
The `discrete' quantum mechanics of the Meixner-Pollaczek,
the continuous Hahn, the continuous dual Hahn, the Wilson and
the Askey-Wilson polynomials discussed in \cite{os4,os5,os6,os7} are
only for restricted parameter ranges; for example the angle was
$\phi=\pi/2$ for the Meixner-Pollaczek polynomial and all the
parameters were restricted real for the  continuous Hahn,
the continuous dual Hahn, the Wilson and the Askey-Wilson polynomials.
This is due to a historical reason that these polynomials with
the restricted parameter ranges were first recognised by the present authors
as describing
the classical equilibrium positions \cite{rags,os3,os4,os5,os6,vD04}
of multi-particle exactly solvable dynamical systems of
Ruijsenaars-Schneider-van Diejen type \cite{RS,vD}.
It is a deformation of the classical results dating as far back as
Stieltjes \cite{Stiel}, \cite{calnuovo,cs1}
that the classical equilibrium positions of multi-particle exactly
solvable dynamical systems of Calogero-Sutherland type \cite{cal,sut}
are described by the zeros of the classical orthogonal polynomials
(the Hermite, Laguerre and Jacobi).
The `discrete' quantum mechanics was constructed \cite{os4,os5,os6}
based on the analogy that these orthogonal polynomials would
constitute the eigenfunctions of certain quantum mechanical systems
in the same way as the classical orthogonal polynomials (the Hermite,
Laguerre and Jacobi) do.
As will be shown in detail in the main text, these orthogonal
polynomials enjoy the exact solvability and related properties for
the full ranges of the parameters.
Attempts to further deform these exactly solvable systems have yielded
several examples \cite{st1,os10,newqes} of the so-called quasi-exactly
solvable systems \cite{Ush,turb}.
Another objective of the present paper is to explore in detail the
properties of the systems obtained by restricting the Askey-Wilson
system, treated in \S\ref{[KS3.3]}--\S\ref{[KS3.9]}.
Some of these have interesting and useful forms of the dynamical
symmetry algebras or the explicit forms of coherent state, etc,
as evidenced by the $q$-oscillator algebras realised by the continuous
(big) $q$-Hermite polynomial \cite{os11}. Aspects of ordinary theory of
orthogonal polynomials are not particularly emphasised.

This paper is organised as follows. In section two, the general setting
of the `discrete' quantum mechanics is recapitulated with appropriate
notation. Starting with the parameters in the potential function and
the Hamiltonian, various concepts and solution methods are briefly
surveyed. Sections three to five are the main body of the paper,
discussing various examples of exactly solvable `discrete' quantum
mechanics. They are divided into three groups according to the
sinusoidal coordinate $\eta(x)$.
Section three is for the polynomials in $\eta(x)=x$.
Section four is for the polynomials in $\eta(x)=x^2$.
Section five is for the polynomials in $\eta(x)=\cos x$.
Very roughly speaking, polynomials in section three are the deformation
of the Hermite polynomial; those in section four are the deformation of
the Laguerre polynomial and those in sections five  are the
deformation of the Jacobi polynomial from the point of view of the
sinusoidal coordinates, but not from the energy spectrum.
Section six is for a summary and comments.
Appendix A provides a diagrammatic proof of the hermiticity
(self-adjointness) of the Hamiltonians of `discrete' quantum mechanics.
Appendix B is a collection of the definition of basic symbols and
functions used in this paper for self-containedness.

\section{General setting}
\label{setting}
\setcounter{equation}{0}

The dynamical variables are the coordinate $x$ ($x\in\mathbb{R}$) and
the conjugate momentum $p$, which is realised as a differential
operator $p=-id/dx$.
The other parameters are symbolically denoted as
$\bm{\lambda}=(\lambda_1,\lambda_2,\ldots)$ on top of $q$ ($0<q<1$)
and $\phi$ ($\phi\in\mathbb{R}$). For the $q$-systems, the parameters
are denoted as $q^{\bm{\lambda}}=(q^{\lambda_1},q^{\lambda_2},\ldots)$.
Complex conjugation is denoted by ${}^*$ and the absolute value
$|f(x)|$ is $|f(x)|=\sqrt{f(x)f(x)^*}$.
Here $f(x)^*$ means $(f(x))^*$ and $f(x)^*|_{x\to x+a}=f(x+a^*)^*$,
since $x$ is real.

\paragraph{Hamiltonian}
\label{praham}
The Hamiltonian has a general form
\begin{equation}
  \mathcal{H}\eqdef\sqrt{V(x)}\,e^{\gamma p}\sqrt{V(x)^*}
  +\sqrt{V(x)^*}\,e^{-\gamma p}\sqrt{V(x)}-V(x)-V(x)^*,
  \label{H}
\end{equation}
in which $\gamma$ is a real constant. It is either $1$ or $\log q$.
The potential function $V$ depends on the parameters,
$V(x)=V(x\,;\bm{\lambda})$, whereas the $q$ and $\phi$ dependence is
not explicitly indicated.
The parameter dependence of the Hamiltonian
$\mathcal{H}=\mathcal{H}(\bm{\lambda})$ is not explicitly indicated
in most cases.

The eigenvalue problem or the time-independent Schr\"odinger equation
is a difference equation in stead of differential in ordinary quantum
mechanics:
\begin{equation}
  \mathcal{H}\phi_n(x)=\mathcal{E}_n\phi_n(x)\quad
  (n=0,1,2,\ldots),\quad
  \mathcal{E}_0<\mathcal{E}_1<\mathcal{E}_2<\cdots,
\end{equation}
in which $\phi_n(x)=\phi_n(x\,;\bm{\lambda})$ is the eigenfunction
belonging to the energy eigenvalue
$\mathcal{E}_n=\mathcal{E}_n(\bm{\lambda})$.
The difference equation has inherent non-uniqueness of solutions;
if $\phi(x)$ is a solution so is $\phi(x)Q(x)$ when $Q(x)$ is any
periodic function with the period $i\gamma$.
This non-uniqueness problem is resolved when the Hilbert space
of the state vectors is specified. See Appendix A.

\paragraph{Factorisation}
\label{parafact}
Factorisation of the Hamiltonian is an important property
\begin{equation}
  \mathcal{H}=T_++T_--V(x)-V(x)^*
  =(S_+^{\dagger}-S_-^{\dagger})(S_+-S_-)
  =\mathcal{A}^{\dagger}\mathcal{A},
  \label{factham}
\end{equation}
in which various quantities $S_{\pm}=S_{\pm}(\bm{\lambda})$,
$T_{\pm}=T_{\pm}(\bm{\lambda})$, $\mathcal{A}=\mathcal{A}(\bm{\lambda})$
are defined as (${}^\dagger$ denote the hermitian conjugation with
respect to the chosen inner product \eqref{inner} and 
\eqref{innera1}--\eqref{innera3}):
\begin{gather}
  \!
  S_+\eqdef e^{\gamma p/2}\sqrt{V(x)^*},\ \,
  S_-\eqdef e^{-\gamma p/2}\sqrt{V(x)},\ \,
  S_+^{\dagger}\eqdef\sqrt{V(x)}\,e^{\gamma p/2},\ \,
  S_-^{\dagger}\eqdef\sqrt{V(x)^*}\,e^{-\gamma p/2},
  \label{spmdef}\\
  T_+\eqdef S_+^{\dagger}S_+=\sqrt{V(x)}\,e^{\gamma p}\sqrt{V(x)^*},\quad
  T_-\eqdef S_-^{\dagger}S_-=\sqrt{V(x)^*}\,e^{-\gamma p}\sqrt{V(x)},
  \label{tpmdef}\\
  \mathcal{A}\eqdef i(S_+-S_-),\quad
  \mathcal{A}^{\dagger}\eqdef-i(S_+^{\dagger}-S_-^{\dagger}).
  \label{aaddef}
\end{gather}

\paragraph{Ground state wavefunction}
\label{paraground}
The ground state wavefunction $\phi_0(x)=\phi_0(x\,;\bm{\lambda})$ is
annihilated by the $\mathcal{A}$ operator
\begin{equation}
  \mathcal{A}\phi_0(x)=0\ \Rightarrow \mathcal{H}\phi_0(x)=0 \ \Rightarrow
  \mathcal{E}_0=0,
  \label{groundeig}
\end{equation}
which is a zero mode of the Hamiltonian. The above equation reads
explicitly as
\begin{equation}
  \sqrt{V(x+\tfrac{i\gamma}{2})^*}\,\phi_0(x-\tfrac{i\gamma}{2})
  =\sqrt{V(x+\tfrac{i\gamma}{2})}\,\phi_0(x+\tfrac{i\gamma}{2}).
  \label{gs1}
\end{equation}
Among possible solutions, we choose a real and nodeless $\phi_0$.
As will be shown in Appendix A, the requirement of the hermiticity
(self-adjointness) of the Hamiltonian $\mathcal{H}$ selects a unique
solution $\phi_0$, which is given explicitly in
each subsection \eqref{Hahnphi0}, \eqref{Meixphi0}, \eqref{Wilsonphi0},
\eqref{dualHahnphi0}, \eqref{AWilsonphi0}, \eqref{qHahnphi0},
\eqref{ASCphi0}, \eqref{bqHphi0}, \eqref{qHphi0}, \eqref{qJphi0} and
\eqref{qLphi0}.

\paragraph{Similarity transformed Hamiltonian}
\label{parasimham}
The similarity transformed Hamiltonian
$\widetilde{\mathcal{H}}=\widetilde{\mathcal{H}}(\bm{\lambda})$
in terms of the ground state wavefunction $\phi_0$ \eqref{gs1} is
\begin{align}
  \widetilde{\mathcal{H}}&\eqdef\phi_0(x)^{-1}\circ\mathcal{H}\circ \phi_0(x)
  =\widetilde{T}_++\widetilde{T}_--V(x)-V(x)^*\n
  &=V(x)\,e^{\gamma p}+V(x)^*\,e^{-\gamma p}-V(x)-V(x)^*,
\end{align}
in which $ \widetilde{T}_\pm$ are defined as
\begin{equation}
  \widetilde{T}_+\eqdef\phi_0(x)^{-1}\circ T_+\circ \phi_0(x)
  =V(x)\,e^{\gamma p},\quad
  \widetilde{T}_-\eqdef\phi_0(x)^{-1}\circ T_-\circ \phi_0(x)
  =V(x)^*\,e^{-\gamma p}.
\end{equation}
It acts on the polynomial part of the eigenfunction.
Let us write the excited state eigenfunction
$\phi_n(x)=\phi_n(x\,;\bm{\lambda})$ as
\begin{equation}
  \phi_n(x\,;\bm{\lambda})
  =\phi_0(x\,;\bm{\lambda})P_n(\eta(x)\,;\bm{\lambda}),
\end{equation}
in which $P_n(\eta)=P_n(\eta\,;\bm{\lambda})$ is a polynomial in the
{\em sinusoidal coordinate\/} $\eta(x)$ \cite{os7}.
Here $\eta(x)$ is a real function of $x$.
The sinusoidal coordinate $\eta(x)$ discussed in this paper has no
$\bm{\lambda}$-dependence in contrast to the cases studied in \cite{os12}.
Then $\widetilde{\mathcal{H}}$ acts on $P_n(\eta)$:
\begin{equation}
  \widetilde{\mathcal{H}}(\bm{\lambda})P_n(\eta(x)\,;\bm{\lambda})
  =\mathcal{E}_n(\bm{\lambda})P_n(\eta(x)\,;\bm{\lambda}).
\end{equation}
For all the examples discussed in this paper, $\widetilde{\mathcal{H}}$
is {\em lower triangular\/} in the special basis
\begin{equation}
  1,\ \eta(x),\ \eta(x)^2,\ \ldots, \eta(x)^n,\ \ldots,
\end{equation}
spanned by the  sinusoidal coordinate $\eta(x)$
($\eta(x)=x, x^2, \cos x$) \cite{os4,os5,os6,os7}:
\begin{equation}
  \widetilde{\mathcal{H}}(\bm{\lambda})\eta(x)^n
  =\mathcal{E}_n(\bm{\lambda})\eta(x)^n+\text{lower orders in}\ \eta(x).
  \label{lowtri}
\end{equation}

\begin{figure}[t]
  \centering
  \includegraphics*[scale=0.9]{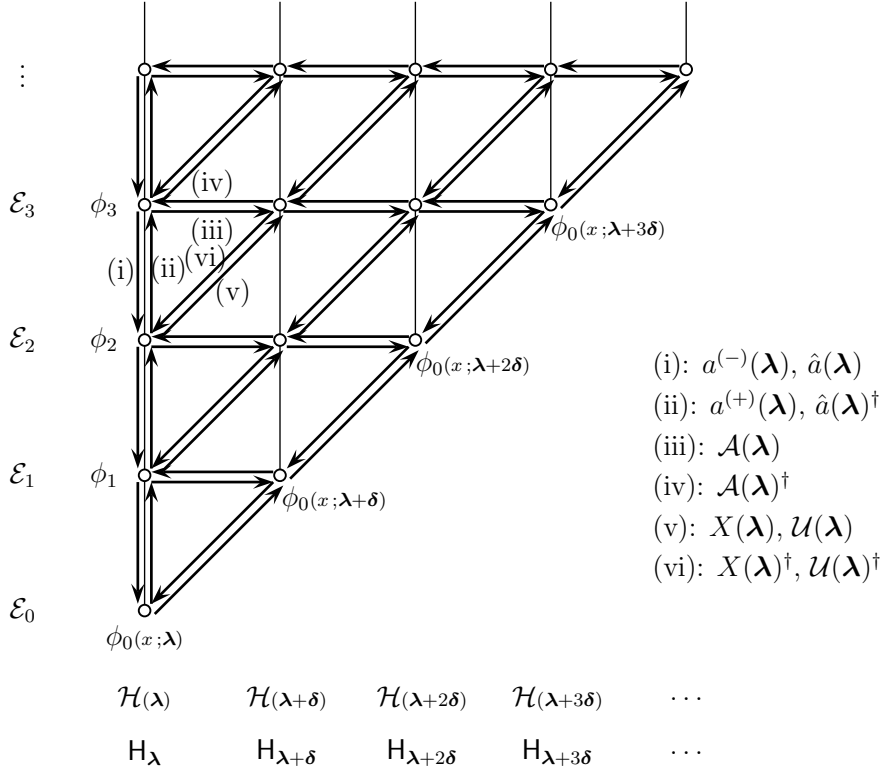}
  \caption{Shape invariance and Crum's scheme.}
  \label{crumfig}
\end{figure}

\paragraph{Shape invariance}
\label{parashape}
The factorised Hamiltonian \eqref{factham} has the dynamical symmetry
called {\em  shape invariance\/} \cite{genden} if the following
relation holds:
\begin{equation}
  \mathcal{A}(\bm{\lambda})\mathcal{A}(\bm{\lambda})^{\dagger}
  =\kappa\mathcal{A}(\bm{\lambda}+\bm{\delta})^{\dagger}
  \mathcal{A}(\bm{\lambda}+\bm{\delta})+\mathcal{E}_1(\bm{\lambda}),
  \label{shapeinv}
\end{equation}
in which $\kappa$ is a real positive parameter and $\bm{\delta}$
denotes the shift of the parameters and $\mathcal{E}_1(\bm{\lambda})$
is the eigenvalue of the first excited state.
This relation is satisfied by all the examples discussed in this paper.
Shape invariance means that the original Hamiltonian
$\mathcal{H}(\bm{\lambda})$ and the {\em associated\/} Hamiltonian
$\mathcal{A}(\bm{\lambda})\mathcal{A}(\bm{\lambda})^{\dagger}$
in Crum's \cite{crum} sense (or the susy partner Hamiltonian
in the so-called supersymmetric quantum mechanics \cite{infhul,susyqm})
have the same {\em shape\/} up to a multiplicative factor $\kappa$
and an additive constant $\mathcal{E}_1(\bm{\lambda})$.
In terms of the potential function $V(x\,;\bm{\lambda})$, the above
relation reads explicitly as
\begin{align}
  &V(x-\tfrac{i\gamma}{2}\,;\bm{\lambda})
  V(x+\tfrac{i\gamma}{2}\,;\bm{\lambda})^*
  =\kappa^2\,V(x\,;\bm{\lambda}+\bm{\delta})
  V(x+i\gamma\,;\bm{\lambda}+\bm{\delta})^*,
  \label{shapeinv1}\\
  &V(x+\tfrac{i\gamma}{2}\,;\bm{\lambda})
  +V(x+\tfrac{i\gamma}{2}\,;\bm{\lambda})^*
  =\kappa\bigl(V(x\,;\bm{\lambda}+\bm{\delta})
  +V(x\,;\bm{\lambda}+\bm{\delta})^*\bigr)
  -\mathcal{E}_1(\bm{\lambda}).
  \label{shapeinv2}
\end{align}
Among many consequences of shape invariance,
we list the most salient ones. All the eigenvalues are generated by
$\mathcal{E}_1(\bm{\lambda})$ and the corresponding eigenfunctions
are generated from the known form of the ground state eigenfunction
$\phi_0$ \eqref{groundeig} together with the multiple action of the
successive $\mathcal{A}^\dagger$ operator \cite{os4,os5,os6}:
\begin{align}
  &\mathcal{E}_n(\bm{\lambda})=\sum_{s=0}^{n-1}
  \kappa^s\mathcal{E}_1(\bm{\lambda}+s\bm{\delta}),\\
  &\phi_n(x\,;\bm{\lambda})\propto
  \mathcal{A}(\bm{\lambda})^{\dagger}
  \mathcal{A}(\bm{\lambda}+\bm{\delta})^{\dagger}
  \mathcal{A}(\bm{\lambda}+2\bm{\delta})^{\dagger}
  \cdots
  \mathcal{A}(\bm{\lambda}+(n-1)\bm{\delta})^{\dagger}
  \phi_0(x\,;\bm{\lambda}+n\bm{\delta}).
  \label{phin=A..Aphi0}
\end{align}
The latter is related to a Rodrigues type formula for the eigenpolynomials.
We illustrate the shape invariance and Crum's scheme in
Fig.\ref{crumfig} at the end of this section. The Hilbert space 
belonging to the Hamiltonian
$\mathcal{H}(\bm{\lambda})$ is denoted as $\mathsf{H}_{\bm{\lambda}}$.

\paragraph{Closure relation}
\label{paraclos}
Another important symmetry concept of exactly solvable quantum mechanics
is the {\em closure relation\/} \cite{os7,os12}:
\begin{equation}
  [\mathcal{H},[\mathcal{H},\eta]\,]=\eta\,R_0(\mathcal{H})
  +[\mathcal{H},\eta]\,R_1(\mathcal{H})+R_{-1}(\mathcal{H}).
  \label{closurerel}
\end{equation}
Here $\eta(x)$ is the sinusoidal coordinate and $R_i(\mathcal{H})$ is
a polynomial in  $\mathcal{H}$.
At the classical mechanics level, it is easy to see that the closure
relation means that $\eta(x)$ undergoes a sinusoidal motion with
frequency $\sqrt{\mathcal{R}_0(\mathcal{E})}$.
The closure relation \eqref{closurerel} is satisfied by all the 
examples discussed
in this paper and the explicit forms of $R_i(\mathcal{H})$,
$i=-1,0,1$ and $\mathcal{E}_n(\bm{\lambda})$ are given in each subsection.
The closure relation \eqref{closurerel} enables us to express any
multiple commutator
\(
  [\mathcal{H},[\mathcal{H},\cdots,[\mathcal{H},\eta(x)]
  \!\cdot\!\cdot\cdot]]
\)
as a linear combination of the operators $\eta(x)$ and
$[\mathcal{H},\eta(x)]$ with coefficients depending on the Hamiltonian
$\mathcal{H}$ only. As we will see shortly, the exact Heisenberg
operator solution and the annihilation/creation operators are obtained
as a consequence \cite{os7,os12}.

Let us consider the closure relation \eqref{closurerel} as an algebraic
constraint on $\eta(x)$ and the Hamiltonian, for given constants
$\{r_i^{(j)}\}$.
The l.h.s. consists of $e^{2\gamma p}$, $e^{\gamma p}$, $1$,
$e^{-\gamma p}$, $e^{-2\gamma p}$ , then $R_i$ can be parametrised as
\begin{equation}
  R_0(y)=r_0^{(2)}y^2+r_0^{(1)}y+r_0^{(0)},\quad
  R_1(y)=r_1^{(1)}y+r_1^{(0)},\quad
  R_{-1}(y)=r_{-1}^{(2)}y^2+r_{-1}^{(1)}y+r_{-1}^{(0)}.
\end{equation}
The similarity transformation of \eqref{closurerel}
\begin{equation}
  [\widetilde{\mathcal{H}},[\widetilde{\mathcal{H}},\eta]\,]
  =\eta\,R_0(\widetilde{\mathcal{H}})
  +[\widetilde{\mathcal{H}},\eta]\,R_1(\widetilde{\mathcal{H}})
  +R_{-1}(\widetilde{\mathcal{H}})
  \label{closurerelt}
\end{equation}
gives rise to the following five conditions:
\begin{align}
  &\eta(x-2i\gamma)-2\eta(x-i\gamma)+\eta(x)=r_0^{(2)}\eta(x)+r_{-1}^{(2)}
  +r_1^{(1)}\bigl(\eta(x-i\gamma)-\eta(x)\bigr),
  \label{closurerel1}\\
  &\eta(x+2i\gamma)-2\eta(x+i\gamma)+\eta(x)=r_0^{(2)}\eta(x)+r_{-1}^{(2)}
  +r_1^{(1)}\bigl(\eta(x+i\gamma)-\eta(x)\bigr),
  \label{closurerel1p}\\
  &\bigl(\eta(x-i\gamma)-\eta(x)\bigr)
  \bigl(V(x-i\gamma)+V(x+i\gamma)^*-V(x)-V(x)^*\bigr)\n
  &\quad=-\bigl(r_0^{(2)}\eta(x)+r_{-1}^{(2)}\bigr)
  \bigl(V(x-i\gamma)+V(x+i\gamma)^*+V(x)+V(x)^*\bigr)\n
  &\quad\phantom{=}
  -r_1^{(1)}\bigl(\eta(x-i\gamma)-\eta(x)\bigr)
  \bigl(V(x-i\gamma)+V(x+i\gamma)^*\bigr)\n
  &\quad\phantom{=}
  +r_0^{(1)}\eta(x)+r_{-1}^{(1)}
  +r_1^{(0)}\bigl(\eta(x-i\gamma)-\eta(x)\bigr),
  \label{closurerel2}\\
  &\bigl(\eta(x+i\gamma)-\eta(x)\bigr)
  \bigl(V(x-i\gamma)^*+V(x+i\gamma)-V(x)^*-V(x)\bigr)\n
  &\quad=-\bigl(r_0^{(2)}\eta(x)+r_{-1}^{(2)}\bigr)
  \bigl(V(x-i\gamma)^*+V(x+i\gamma)+V(x)^*+V(x)\bigr)\n
  &\quad\phantom{=}
  -r_1^{(1)}\bigl(\eta(x+i\gamma)-\eta(x)\bigr)
  \bigl(V(x-i\gamma)^*+V(x+i\gamma)\bigr)\n
  &\quad\phantom{=}
  +r_0^{(1)}\eta(x)+r_{-1}^{(1)}
  +r_1^{(0)}\bigl(\eta(x+i\gamma)-\eta(x)\bigr),
  \label{closurerel2p}\\
  &2\bigl(\eta(x)-\eta(x-i\gamma)\bigr)V(x)V(x+i\gamma)^*
  +2\bigl(\eta(x)-\eta(x+i\gamma)\bigr)V(x)^*V(x+i\gamma)\n
  &\quad=\bigl(r_0^{(2)}\eta(x)+r_{-1}^{(2)}\bigr)
  \bigl(V(x)V(x+i\gamma)^*+V(x)^*V(x+i\gamma)+\bigl(V(x)+V(x)^*\bigr)^2
  \bigr)\n
  &\quad\phantom{=}
  +r_1^{(1)}\bigl(\eta(x-i\gamma)-\eta(x)\bigr)V(x)V(x+i\gamma)^*
  +r_1^{(1)}\bigl(\eta(x+i\gamma)-\eta(x)\bigr)V(x)^*V(x+i\gamma)\n
  &\quad\phantom{=}
  -\bigl(r_0^{(1)}\eta(x)+r_{-1}^{(1)}\bigr)\bigl(V(x)+V(x)^*\bigr)
  +r_0^{(0)}\eta(x)+r_{-1}^{(0)}.
  \label{closurerel3}
\end{align}
For real $\{r_i^{(j)}\}$ (this is indeed the case for all the examples
discussed in this paper), \eqref{closurerel1p} and \eqref{closurerel2p}
are the complex conjugate of \eqref{closurerel1} and \eqref{closurerel2},
respectively.

In contrast to the cases of the orthogonal polynomials with discrete
measures discussed in section 4 of \cite{os12}, the determination of
$\eta(x)$ and the possible forms of $V(x)$ is not straightforward
due to the ambiguities of periodic functions with $i\gamma$ period.
Here we mention only the basic results.
It is easy to see that \eqref{closurerel1}--\eqref{closurerel2p}
require $r_0^{(2)}=r_1^{(1)}$ and $r_0^{(1)}=2r_1^{(0)}$, which is
consistent with the hermitian conjugation of \eqref{closurerel}.
With these constraints, the first condition \eqref{closurerel1} reads
with $x\to x+i\gamma$
\begin{equation}
  \eta(x-i\gamma)-(2+r_1^{(1)})\eta(x)+\eta(x+i\gamma)=r_{-1}^{(2)}.
  \label{crcond1pp}
\end{equation}
Following the arguments given in section 4 and appendix A of \cite{os12},
we deduce from \eqref{closurerel2} and \eqref{closurerel3} the general
relationship
\begin{align}
  &\bigl(\eta(x-i\gamma)-\eta(x)\bigr)\bigl(\eta(x+i\gamma)-\eta(x)\bigr)
  \left(V(x)+V(x)^*\right)\n
  &\ \ =
  -r_1^{(0)}\eta(x)^2-r_{-1}^{(1)}\eta(x)-C_1(x),
  \label{ansa}\\
  &\bigl(\eta(x-2i\gamma)-\eta(x)\bigr)
  \bigl(\eta(x-i\gamma)-\eta(x+i\gamma)\bigr)V(x)V(x+i\gamma)^*\n
  &\ \ =
  \frac{\bigl(r_1^{(0)}\eta(x-i\gamma)\eta(x)
             +r_{-1}^{(1)}\eta(x-i\gamma)+C_1(x)\bigr)
        \bigl(r_1^{(0)}\eta(x-i\gamma)\eta(x)+r_{-1}^{(1)}\eta(x)+C_1(x)\bigr)}
       {\bigl(\eta(x-i\gamma)-\eta(x)\bigr)^2}\n
  &\ \ \phantom{=}
  -r_0^{(0)}\eta(x-i\gamma)\eta(x)
  -r_{-1}^{(0)}\bigl(\eta(x-i\gamma)+\eta(x)\bigr)
  +C_2(x),
  \label{ansVV^*}
\end{align}
in which $C_j(x)$ ($j=1,2$) is an arbitrary function satisfying
the periodicity $C_j(x+i\gamma)=C_j(x)$.
The hermiticity of the Hamiltonian $\mathcal{H}$ would restrict $C_j(x)$
severely.
Further analysis of the closure relation
\eqref{closurerel1}--\eqref{closurerel3} will be published elsewhere.

Like the cases of discrete measures \cite{os12}, the dual closure
relation
\begin{equation}
  [\eta,[\eta,\mathcal{H}]\,]=\mathcal{H}\,R_0^{\text{dual}}(\eta)
  +[\eta,\mathcal{H}]\,R_1^{\text{dual}}(\eta)+R_{-1}^{\text{dual}}(\eta)
  \label{dualclosurerel}
\end{equation}
holds and $R_i^{\text{dual}}$ are given by
\begin{align}
  R_1^{\text{dual}}(\eta(x))&=
  \bigl(\eta(x-i\gamma)-\eta(x)\bigr)+\bigl(\eta(x+i\gamma)-\eta(x)\bigr),
  \label{dualclcon1}\\
  R_0^{\text{dual}}(\eta(x))&=
  -\bigl(\eta(x-i\gamma)-\eta(x)\bigr)\bigl(\eta(x+i\gamma)-\eta(x)\bigr),\\
  R_{-1}^{\text{dual}}(\eta(x))&
  =\bigl(V(x)+V(x)^*\bigr)R_0^{\text{dual}}(\eta(x)).
  \label{dualclcon3}
\end{align}
Eqs. \eqref{crcond1pp} and \eqref{ansa} imply
$R_1^{\text{dual}}(y)=r_1^{(1)}y+r_{-1}^{(2)}$ and
$R_{-1}^{\text{dual}}(\eta(x))=r_1^{(0)}\eta(x)^2+r_{-1}^{(1)}\eta(x)
+C_1(x)$.

\paragraph{Auxiliary function $\varphi$}
\label{paraaux}
In all the examples discussed in this paper, the ground state
wavefunction with shifted $x$ and parameters
$\phi_0(x-\tfrac{i\gamma}{2}\,;\bm{\lambda}+\bm{\delta})$ is related
to its original value $\phi_0(x\,;\bm{\lambda})$ via a real auxiliary
function $\varphi$:
\begin{equation}
  \phi_0(x-\tfrac{i\gamma}{2}\,;\bm{\lambda}+\bm{\delta})
  =\sqrt{V(x\,;\bm{\lambda})}\,\varphi(x-\tfrac{i\gamma}{2})
  \phi_0(x\,;\bm{\lambda}).
  \label{gs2}
\end{equation}
The auxiliary function $\varphi(x)$ discussed in this paper has no
$\bm{\lambda}$-dependence in contrast to the cases studied in \cite{os12}.
It is easy to see that \eqref{gs2} implies \eqref{gs1}.
The explicit forms of $\varphi(x)$ are given at
the beginning of each section \eqref{first3}, \eqref{first4}, \eqref{first5}.

\paragraph{`Similarity' transformation II}
`Similarity' transformed Hamiltonian or that of
$S_\pm$, $S_\pm^\dagger$
operators \eqref{spmdef} take simpler forms with the help of the
auxiliary function $\varphi$ \eqref{gs2}:
\begin{align}
  \phi_0(x\,;\bm{\lambda}+\bm{\delta})^{-1}\circ S_{\pm}(\bm{\lambda})\circ
  \phi_0(x\,;\bm{\lambda})&=\varphi(x)^{-1}\,e^{\pm\gamma p/2},
  \label{pSp}\\
  \phi_0(x\,;\bm{\lambda})^{-1}\circ S_{\pm}(\bm{\lambda})^{\dagger}\circ
  \phi_0(x\,;\bm{\lambda}+\bm{\delta})&=
  \begin{cases}
    V(x\,;\bm{\lambda})\,e^{\gamma p/2}\,\varphi(x)\,,\\
    V(x\,;\bm{\lambda})^*\,e^{-\gamma p/2}\,\varphi(x)\,.
  \end{cases}
  \label{pSdp}
\end{align}
Note that the parameter shifts $\pm\bm{\delta}$ are properly incorporated.

\paragraph{Forward/Backward shift operators}
With \eqref{pSp}--\eqref{pSdp} the `similarity' transformed
$\mathcal{A}$ and $\mathcal{A}^\dagger$ operators are obtained.
They are called the forward/backward shift operators:
\begin{align}
  \widetilde{\mathcal{H}}(\bm{\lambda})
  &=\mathcal{B}(\bm{\lambda})\mathcal{F}(\bm{\lambda}),\\
  \mathcal{F}(\bm{\lambda})&\eqdef
  \phi_0(x\,;\bm{\lambda}+\bm{\delta})^{-1}\circ\mathcal{A}(\bm{\lambda})\circ
  \phi_0(x\,;\bm{\lambda})=i\,\varphi(x)^{-1}
  \bigl(e^{\gamma p/2}-e^{-\gamma p/2}\bigr),
  \label{Fshift}\\
  \mathcal{B}(\bm{\lambda})&\eqdef
  \phi_0(x\,;\bm{\lambda})^{-1}\circ\mathcal{A}(\bm{\lambda})^{\dagger}\circ
  \phi_0(x\,;\bm{\lambda}+\bm{\delta})
  =-i\bigl(V(x\,;\bm{\lambda})\,e^{\gamma p/2}
  -V(x\,;\bm{\lambda})^*\,e^{-\gamma p/2}\bigr)
  \varphi(x).
  \label{Bshift}
\end{align}
The action of the forward shift operator $\mathcal{F}(\bm{\lambda})$
and the backward shift operator $\mathcal{B}(\bm{\lambda})$ on the
polynomial $P_n(\eta\,;\bm{\lambda})$ are:
\begin{align}
  \mathcal{F}(\bm{\lambda})P_n(\eta\,;\bm{\lambda})
  &=f_n(\bm{\lambda})P_{n-1}(\eta\,;\bm{\lambda}+\bm{\delta}),
  \label{FPn}\\
  \mathcal{B}(\bm{\lambda})P_n(\eta\,;\bm{\lambda}+\bm{\delta})
  &=b_n(\bm{\lambda})P_{n+1}(\eta\,;\bm{\lambda}),
  \label{BPn}
\end{align}
in which $f_n(\bm{\lambda})$ and $b_n(\bm{\lambda})$ are real constants
related to $\mathcal{E}_n(\bm{\lambda})$:
\begin{equation}
  f_n(\bm{\lambda})b_{n-1}(\bm{\lambda})=\mathcal{E}_n(\bm{\lambda}).
\end{equation}
For the cases studied in \cite{os12} $b_n(\bm{\lambda})$ is actually
independent of $n$, but here it depends on $n$.
In terms of the forward and backward shift operators, the shape
invariance condition \eqref{shapeinv} reads
\begin{equation}
  \mathcal{F}(\bm{\lambda})\mathcal{B}(\bm{\lambda})
  =\kappa\mathcal{B}(\bm{\lambda}+\bm{\delta})
  \mathcal{F}(\bm{\lambda}+\bm{\delta})
  +\mathcal{E}_1(\bm{\lambda}).
\end{equation}
Corresponding to \eqref{phin=A..Aphi0}, a Rodrigues type formula for
the eigenpolynomials is
\begin{equation}
  P_n(\eta\,;\bm{\lambda})
  =\frac{\mathcal{B}(\bm{\lambda})}{b_{n-1}(\bm{\lambda})}\,
  \frac{\mathcal{B}(\bm{\lambda}+\bm{\delta})}
  {b_{n-2}(\bm{\lambda}+\bm{\delta})}\,
  \frac{\mathcal{B}(\bm{\lambda}+2\bm{\delta})}
  {b_{n-3}(\bm{\lambda}+2\bm{\delta})}
  \cdots
  \frac{\mathcal{B}(\bm{\lambda}+(n-1)\bm{\delta})}
  {b_0(\bm{\lambda}+(n-1)\bm{\delta})}
  \cdot P_0(\eta\,;\bm{\lambda}+n\bm{\delta}),
\end{equation}
where $P_0(\eta\,;\bm{\lambda}+n\bm{\delta})=1$ for all the examples
given in this paper.
With these quantities the action of $\mathcal{A}(\bm{\lambda})$ and
$\mathcal{A}(\bm{\lambda})^{\dagger}$ on the eigenfunction $\phi_n$
can be simply expressed as
\begin{align}
  \mathcal{A}(\bm{\lambda})\phi_n(x\,;\bm{\lambda})
  &=f_n(\bm{\lambda})\phi_{n-1}(x\,;\bm{\lambda}+\bm{\delta}),
  \label{Aphin}\\
  \mathcal{A}(\bm{\lambda})^{\dagger}\phi_n(x\,;\bm{\lambda}+\bm{\delta})
  &=b_n(\bm{\lambda})\phi_{n+1}(x\,;\bm{\lambda}).
  \label{Adphin}
\end{align}

\paragraph{Three term recurrence relation}
\label{parathree}
The polynomial part of the eigenfunction $P_n(\eta)$ is an orthogonal
polynomial with the measure $\phi_0(x)^2$. It satisfies three term
recurrence relations \cite{askey,ismail}.
Let us first write the relation for the monic polynomial
$P_n^{\text{monic}}(\eta)=\eta^n+$ lower degree in $\eta$:
\begin{align}
  &P_n(\eta)=c_nP_n^{\text{monic}}(\eta),\\
  &P_{n+1}^{\text{monic}}(\eta)-(\eta-a^{\text{rec}}_n)P_n^{\text{monic}}(\eta)
  +b^{\text{rec}}_nP_{n-1}^{\text{monic}}(\eta)=0\quad(n\geq 0),
\end{align}
with $P_{-1}^{\text{monic}}(\eta)=0$.
For $P_n(\eta)$ it reads
\begin{gather}
  \eta P_n(\eta)=A_nP_{n+1}(\eta)+B_nP_n(\eta)+C_nP_{n-1}(\eta),
  \label{threeterm}\\
  A_n=\frac{c_n}{c_{n+1}},\quad
  B_n=a_n^{\text{rec}},\quad
  C_n=\frac{c_n}{c_{n-1}}b_n^{\text{rec}}.
  \label{threetermcoeff}
\end{gather}
Sometimes we write the parameter dependence explicitly as
$P_n(\eta)=P_n(\eta\,;\bm{\lambda})$,
$a^{\text{rec}}_n=a^{\text{rec}}_n(\bm{\lambda})$,
$b^{\text{rec}}_n=b^{\text{rec}}_n(\bm{\lambda})$,
$c_n=c_n(\bm{\lambda})$, $A_n=A_n(\bm{\lambda})$,
$B_n=B_n(\bm{\lambda})$, $C_n=C_n(\bm{\lambda})$,
$f_n(\bm{\lambda})$ and $b_n(\bm{\lambda})$.
They are given in each subsection.

\paragraph{Heisenberg operator and Annihilation-Creation operators}
The exact Heisenberg operator solution for $\eta(x)$ is easily
obtained \cite{os7} from the closure relation \eqref{closurerel}:
\begin{align}
  &e^{it\mathcal{H}}\eta(x)e^{-it\mathcal{H}}
  =a^{(+)}e^{i\alpha_+(\mathcal{H})t}+a^{(-)}e^{i\alpha_-(\mathcal{H})t}
  -R_{-1}(\mathcal{H})R_0(\mathcal{H})^{-1},\\
  &\alpha_{\pm}(\mathcal{H})\eqdef\tfrac12\bigl(R_1(\mathcal{H})
  \pm\sqrt{R_1(\mathcal{H})^2+4R_0(\mathcal{H})}\,\bigr),\\
  &\qquad\qquad
  R_1(\mathcal{H})=\alpha_+(\mathcal{H})+\alpha_-(\mathcal{H}),\quad
  R_0(\mathcal{H})=-\alpha_+(\mathcal{H})\alpha_-(\mathcal{H}),\\
  &a^{(\pm)}\eqdef\pm\Bigl([\mathcal{H},\eta(x)]-\bigl(\eta(x)
  +R_{-1}(\mathcal{H})R_0(\mathcal{H})^{-1}\bigr)\alpha_{\mp}(\mathcal{H})
  \Bigr)
  \bigl(\alpha_+(\mathcal{H})-\alpha_-(\mathcal{H})\bigr)^{-1}
  \label{a^{(pm)}}\\
  &\phantom{a^{(\pm)}}=
  \pm\bigl(\alpha_+(\mathcal{H})-\alpha_-(\mathcal{H})\bigr)^{-1}
  \Bigl([\mathcal{H},\eta(x)]+\alpha_{\pm}(\mathcal{H})\bigl(\eta(x)
  +R_{-1}(\mathcal{H})R_0(\mathcal{H})^{-1}\bigr)\Bigr).
\end{align}
The positive/negative frequency parts of the Heisenberg operator solution,
$a^{(\pm)}$ are the annihilation creation operators
\begin{equation}
  a^{(+)\,\dagger}=a^{(-)},\quad
  a^{(+)}\phi_n(x)=A_n\phi_{n+1}(x),\quad
  a^{(-)}\phi_n(x)=C_n\phi_{n-1}(x).
  \label{apmphi}
\end{equation}
Since
\begin{equation}
  \alpha_{\pm}(\mathcal{E}_n)=\mathcal{E}_{n\pm 1}-\mathcal{E}_n,
  \label{alphapmE}
\end{equation}
we obtain
\begin{equation}
  a^{(\pm)}\phi_n(x)=\frac{\pm1}{\mathcal{E}_{n+1}-\mathcal{E}_{n-1}}
  \Bigl([\mathcal{H},\eta(x)]+(\mathcal{E}_n-\mathcal{E}_{n\mp 1})\eta(x)
  +\frac{R_{-1}(\mathcal{E}_n)}{\mathcal{E}_{n\pm 1}-\mathcal{E}_n}\Bigr)
  \phi_n(x).
\end{equation}

\paragraph{Commutation relations of $a^{(\pm)}$ and $\mathcal{H}$}
Simple commutation relations
\begin{equation}
  [\mathcal{H},a^{(\pm)}]=a^{(\pm)}\alpha_{\pm}(\mathcal{H})
  \label{[H,apm]}
\end{equation}
follow from \eqref{a^{(pm)}} and \eqref{closurerel}.
When applied to $\phi_n$, we obtain with the help of \eqref{alphapmE},
\begin{equation}
  [\mathcal{H},a^{(\pm)}]\phi_n
  =(\mathcal{E}_{n\pm 1}-\mathcal{E}_n)a^{(\pm)}\phi_n.
  \label{[H,apm]phin}
\end{equation}
Commutation relations of $a^{(\pm)}$ are expressed in terms of the
coefficients of the three term recurrence relation by \eqref{apmphi}:
\begin{align}
  &a^{(-)}a^{(+)}\phi_n=A_nC_{n+1}\phi_n=b^{\text{rec}}_{n+1}\phi_n,\quad
  a^{(+)}a^{(-)}\phi_n=C_nA_{n-1}\phi_n=b^{\text{rec}}_{n}\phi_n,\\
  &\Rightarrow\quad
  [a^{(-)},a^{(+)}]\phi_n=(b^{\text{rec}}_{n+1}-b^{\text{rec}}_{n})\phi_n.
  \label{[a-,a+]}
\end{align}
These relation simply mean the operator relations
\begin{gather}
  a^{(-)}a^{(+)}=f(\mathcal{H}),
  \label{amapf}\\
  a^{(+)}a^{(-)}=g(\mathcal{H}),
  \label{apamg}
\end{gather}
in which $f$ and $g$ are analytic functions of $\mathcal{H}$.
In other words, $\mathcal{H}$ and $a^{(\pm)}$ form a
so-called quasi-linear algebra \cite{vinzhed}.
This is because the definition of the annihilation/creation
operators depend only on the closure relation \eqref{closurerel},
without any other inputs.
The situation is quite different from those of the wide variety
of proposed annihilation/creation operators for various
quantum systems \cite{coherents}, most of which were introduced within
the framework of `algebraic theory of coherent states'.
In all these cases there is no guarantee for symmetry relations like
\eqref{amapf}, \eqref{apamg}.

In many cases it is convenient to introduce the `number operator'
(or the `level operator') $\mathcal{N}$
\begin{equation}
  \mathcal{N}\phi_n\eqdef n\phi_n.
\end{equation}
For the following types of energy spectra, the number operator
$\mathcal{N}$ can be expressed as a function of the Hamiltonian
$\mathcal{H}$:
\begin{alignat}{2}
  \mathcal{E}_n&=an\ \ (a>0)&\Rightarrow\ \ \mathcal{N}&=a^{-1}\mathcal{H},\\
  \mathcal{E}_n&=n(n+b)\ \ (b>0)&\Rightarrow\ \ \mathcal{N}
  &=\sqrt{\mathcal{H}+\tfrac14b^2}-\tfrac12b,\\
  \mathcal{E}_n&=q^{-n}-1\ \ &\Rightarrow
  \ q^{\mathcal{N}}&=(\mathcal{H}+1)^{-1},\\
  \mathcal{E}_n&=(q^{-n}-1)(1-bq^n)\ \ (0<b<1)\ \ &\Rightarrow
  \ q^{\mathcal{N}}
  &=\frac{1}{2b}\bigl(\mathcal{H}+b+1-\sqrt{(\mathcal{H}+b+1)^2-4b}\,\bigr).
\end{alignat}
Obviously the Hamiltonian is expressed as
$\mathcal{H}=\mathcal{E}_{\mathcal{N}}$.
Then \eqref{[a-,a+]} can be expressed simply as
\begin{equation}
  [a^{(-)},a^{(+)}]=b^{\text{rec}}_{\mathcal{N}+1}
  -b^{\text{rec}}_{\mathcal{N}}
\end{equation}
and \eqref{[H,apm]phin} is rewritten as
\begin{equation}
  [\mathcal{H},a^{(\pm)}]
  =\mathcal{E}_{\mathcal{N}}a^{(\pm)}-a^{(\pm)}\mathcal{E}_{\mathcal{N}}
  =a^{(\pm)}(\mathcal{E}_{\mathcal{N}\pm 1}-\mathcal{E}_{\mathcal{N}}).
\end{equation}
With a deformed commutator
\begin{equation}
  [A,B]_{\alpha}\eqdef AB-\alpha BA,
\end{equation}
we have
\begin{equation}
  [a^{(-)},a^{(+)}]_{\alpha}=b^{\text{rec}}_{\mathcal{N}+1}
  -\alpha b^{\text{rec}}_{\mathcal{N}}.
\end{equation}

\paragraph{Orthogonality and normalisation}
\label{paraortho}
The scalar product for the elements of the Hilbert space belonging to
the Hamiltonian $\mathcal{H}$ is
\begin{equation}
  (g,f)\eqdef\int dx\,g(x)^*f(x),
  \label{inner}
\end{equation}
in which the integration range depends on the specific Hamiltonian or
the polynomial.
The orthogonality of the eigenvectors $\{\phi_n(x)\}$,
$\phi_n(x)=\phi_0(x)P_n(\eta(x))$ is:
\begin{equation}
  (\phi_n,\phi_m)=\int dx\,\phi_0(x\,;\bm{\lambda})^2
  P_n(\eta(x)\,;\bm{\lambda})^*P_m(\eta(x)\,;\bm{\lambda})
  =h_n(\bm{\lambda})\delta_{nm},
\end{equation}
in which $h_n(\bm{\lambda})>0$.
The constants $h_n$, $c_n$ and $b_n^{\text{rec}}$ are related as
\begin{equation}
  b^{\text{rec}}_n=\frac{c_{n-1}^2}{c_n^2}\frac{h_n}{h_{n-1}}\quad(n\geq 1),
  \qquad
  h_n=h_0c_n^2\prod_{j=1}^nb^{\text{rec}}_n\quad(n\geq 0).
\end{equation}
Let us denote the $n$-th normalised eigenfunction as
\begin{equation}
   \hat{\phi}_n(x\,;\bm{\lambda})=
   N_n(\bm{\lambda})P_n(\eta(x)\,;\bm{\lambda})
   \hat{\phi}_0(x\,;\bm{\lambda}),\quad
   \hat{\phi}_0(x\,;\bm{\lambda})=
   \frac{\phi_0(x\,;\bm{\lambda})}{\sqrt{h_0(\bm{\lambda})}},\quad
   N_n(\bm{\lambda})=\sqrt{\frac{h_0(\bm{\lambda})}{h_n(\bm{\lambda})}}.
\end{equation}
These normalisation constants are given for each polynomial.

\paragraph{Coherent states}
\label{paracoherent}
There are many different nonequivalent definitions of coherent states.
Here we adopt the most conventional one, as the eigenvector of the
annihilation operator $a^{(-)}$, \eqref{apmphi}:
\begin{equation}
  a^{(-)}\psi(\alpha,x)=\alpha\psi(\alpha,x),
  \quad \alpha\in\mathbb{C}.
\end{equation}
It is expressed in terms of the coefficient $C_n$ of the
three term recurrence relation \eqref{threeterm} and
\eqref{threetermcoeff} as \cite{os7}
\begin{equation}
  \psi(\alpha,x)=\psi(\alpha,x\,;\bm{\lambda})
  =\phi_0(x\,;\bm{\lambda})\sum_{n=0}^{\infty}
  \frac{\alpha^n}{\prod_{k=1}^nC_k}\,P_n(\eta(x)\,;\bm{\lambda}).
  \label{psi}
\end{equation}
Thus we obtain one new coherent state for each polynomial;
\eqref{31coh}, \eqref{32coh}, \eqref{41coh}, \eqref{42coh}, \eqref{51coh},
\eqref{52coh}, \eqref{alsalamcohe}, \eqref{54coh}, \eqref{55coh},
\eqref{56coh} and \eqref{57coh}.
If the sum on the r.h.s. is expressed by a simple function, it is
a generating function of the polynomial $P_n(\eta)$.
In most explicit examples to be discussed in later sections,
the potential functions, the Hamiltonians and thus the polynomials
themselves are totally symmetric in the parameters, see for example,
the Askey-Wilson polynomial \S\ref{[KS3.1]}.
The above coherent state, being totally symmetric, gives the best
candidate for a symmetric generating function.
For the polynomials to be discussed in later sections, however,
most of the known generating functions are not totally symmetric.

\paragraph{$\bm{\lambda}$-shift operators}
Let us fix an orthonormal basis $\{\hat{\phi}_n(x;\bm{\lambda})\}$
and define a unitary operator $\mathcal{U}$ ($\mathcal{U}^\dagger$)
as
\begin{equation}
  \mathcal{U}\hat{\phi}_n(x\,;\bm{\lambda})\eqdef
  \hat{\phi}_{n}(x\,;\bm{\lambda}+\bm{\delta}),\quad
  \mathcal{U}^{\dagger}\hat{\phi}_{n}(x\,;\bm{\lambda}+\bm{\delta})
  =\hat{\phi}_n(x\,;\bm{\lambda}).
\end{equation}
Then we can define another set of annihilation-creation operators
$\hat{a}$, $\hat{a}^{\dagger}$:
\begin{equation}
  \hat{a}\eqdef\mathcal{U}^{\dagger}\mathcal{A},\quad
  \hat{a}^{\dagger}=\mathcal{A}^{\dagger}\,\mathcal{U}.
\end{equation}
They satisfy $\mathcal{H}=\hat{a}^{\dagger}\hat{a}$ and their action
on $\phi_n$ are derived from \eqref{Aphin} and \eqref{Adphin},
$\hat{a}\phi_n(x\,;\bm{\lambda})\propto\phi_{n-1}(x\,;\bm{\lambda})$,
$\hat{a}^{\dagger}\phi_n(x\,;\bm{\lambda})\propto
\phi_{n+1}(x\,;\bm{\lambda})$.
Although this kind of annihilation-creation operators have been
considered in many literature \cite{coherents}, it should be stressed
that they are formal because $\mathcal{U}$ and $\mathcal{U}^{\dagger}$
are formal operators.
On the other hand, $a^{(\pm)}$ obtained from the Heisenberg solution
are explicitly expressed in terms of difference operators
(differential operators, in ordinary quantum mechanics), \eqref{a^{(pm)}}.
Note that the construction method of $\hat{a}$ and $\hat{a}^{\dagger}$
is based on the shape invariance but that of $a^{(\pm)}$ is not.
The latter is based on the closure relation.

The key point of the construction of $\hat{a}$ and $\hat{a}^{\dagger}$
is the proper shift of the parameters $\bm{\lambda}$, which is achieved by
the formal operators $\mathcal{U}$ and $\mathcal{U}^{\dagger}$.
We introduce another set of  $\bm{\lambda}$-shift operators
$X$ and $X^\dagger$ explicitly in terms of
difference operators  through the following relations:
\begin{equation}
  a^{(+)}=\mathcal{A}^{\dagger}X,\quad
  a^{(-)}=X^{\dagger}\mathcal{A}.
\end{equation}
By using the shape invariance \eqref{shapeinv}, we have
\begin{equation}
  \mathcal{A}a^{(+)}=\mathcal{A}\mathcal{A}^{\dagger}X
  =\bigl(\kappa\mathcal{A}(\bm{\lambda}+\bm{\delta})^{\dagger}
  \mathcal{A}(\bm{\lambda}+\bm{\delta})+\mathcal{E}_1\bigr)X
  =\bigl(\kappa\mathcal{H}(\bm{\lambda}+\bm{\delta})
  +\mathcal{E}_1\bigr)X.
\end{equation}
Since $\kappa\mathcal{H}(\bm{\lambda}+\bm{\delta})+\mathcal{E}_1$ is
a positive operator, we obtain
\begin{align}
  X&=\bigl(\kappa\mathcal{H}(\bm{\lambda}+\bm{\delta})
  +\mathcal{E}_1\bigr)^{-1}\mathcal{A}\,a^{(+)}\n
  &=\bigl(\kappa\mathcal{H}(\bm{\lambda}+\bm{\delta})
  +\mathcal{E}_1\bigr)^{-1}\mathcal{A}\n
  &\qquad\times
  \Bigl([\mathcal{H},\eta(x)]-\bigl(\eta(x)
  +R_{-1}(\mathcal{H})R_0(\mathcal{H})^{-1}\bigr)\alpha_{-}(\mathcal{H})
  \Bigr)
  \bigl(\alpha_+(\mathcal{H})-\alpha_-(\mathcal{H})\bigr)^{-1}.
  \label{X2}
\end{align}
Similarly $X^{\dagger}$ is expressed as
\begin{equation}
  X^{\dagger}=a^{(-)}\mathcal{A}^{\dagger}
  \bigl(\kappa\mathcal{H}(\bm{\lambda}+\bm{\delta})
  +\mathcal{E}_1\bigr)^{-1}.
\end{equation}
Their action on $\phi_n$ are
\begin{align}
  X\phi_n(x\,;\bm{\lambda})&=\frac{A_n(\bm{\lambda})}{b_n(\bm{\lambda})}
  \phi_n(x\,;\bm{\lambda}+\bm{\delta}),
  \label{Xphin}\\
  X^{\dagger}\phi_n(x\,;\bm{\lambda}+\bm{\delta})
  &=\frac{C_{n+1}(\bm{\lambda})}{f_{n+1}(\bm{\lambda})}
  \phi_n(x\,;\bm{\lambda}),
  \label{Xdphin}
\end{align}
and the  $\bm{\lambda}$-shift without changing the level $n$ is 
achieved, as expected.
The $\bm{\lambda}$-shift operators for the polynomials
$P_n(\eta(x)\,;\bm{\lambda})$ are given by
$\phi_0(x\,;\bm{\lambda}+\bm{\delta})^{-1}\circ X\circ
\phi_0(x\,;\bm{\lambda})$ and
$\phi_0(x\,;\bm{\lambda})^{-1}\circ X^{\dagger}\circ
\phi_0(x\,;\bm{\lambda}+\bm{\delta})$.
The expression of $X$ and $X^{\dagger}$ may be simplified for some
particular cases, see \S\ref{[KS1.7]}, \S\ref{[KS1.3]}, \S\ref{[KS3.26]}.

Finally we illustrate the shape invariance and Crum's scheme in
Fig.\ref{crumfig}. The Hilbert space belonging to the Hamiltonian
$\mathcal{H}(\bm{\lambda})$ is denoted as $\mathsf{H}_{\bm{\lambda}}$.
The action of various operators and their domains and images are
also illustrated
in Fig.\ref{crumfig}:
\begin{align}
  \mathcal{H}(\bm{\lambda}),\ a^{(\pm)}(\bm{\lambda}),
  \ \hat{a}(\bm{\lambda}), \hat{a}(\bm{\lambda})^{\dagger}
  &:
  \mathsf{H}_{\bm{\lambda}}\to\mathsf{H}_{\bm{\lambda}},\\
  \mathcal{A}(\bm{\lambda}),\ \ X(\bm{\lambda}),
  \ \ \mathcal{U}(\bm{\lambda})
  \ &:
  \mathsf{H}_{\bm{\lambda}}\to\mathsf{H}_{\bm{\lambda}+\bm{\delta}},\\
  \mathcal{A}(\bm{\lambda})^{\dagger},\ X(\bm{\lambda})^{\dagger},
  \ \mathcal{U}(\bm{\lambda})^{\dagger}
  &:
  \mathsf{H}_{\bm{\lambda}+\bm{\delta}}\to\mathsf{H}_{\bm{\lambda}}.
\end{align}

\section{$\eta(x)=x$}
\label{secx}
\setcounter{equation}{0}

{}From this section to section \ref{seccosx}, we present various
formulas and results specific to each example of the exactly solvable
`discrete' quantum mechanics.
These examples are divided into three groups according to the form
of the sinusoidal coordinate; $\eta(x)=x$ in this section,
$\eta(x)=x^2$ in section \ref{secx2}, $\eta(x)=\cos x$ in section
\ref{seccosx}. 
The names of the subsections are taken from the name of the
corresponding orthogonal polynomial and the number, for example,
[KS1.4] indicates the corresponding subsection of the review of
Koekoek and Swarttouw \cite{koeswart}.

In all the examples in this section, we have
\begin{equation}
  \eta(x)=x,\quad -\infty<x<\infty,\quad \gamma=1,
  \quad \kappa=1, \quad \varphi(x)=1.
  \label{first3}
\end{equation}

\subsection{continuous Hahn [KS1.4]}
\label{[KS1.4]}

In previous works \cite{os4,os5,os6,os7}, the parameters $a_1$ and
$a_2$ were restricted to real, positive values. Now they are complex
with positive real parts.

\paragraph{parameters and potential functions}
\begin{equation}
  \bm{\lambda}\eqdef(a_1,a_2),\quad
  \bm{\delta}=(\tfrac12,\tfrac12);\quad \text{Re}\,{a_i}>0;\quad
  V(x\,;\bm{\lambda})\eqdef(a_1+ix)(a_2+ix).
\end{equation}

\paragraph{shape invariance and closure relation}
\begin{align}
  \mathcal{E}_n(\bm{\lambda})&=n(n+b_1-1), \\
  R_1(y)&=2,\quad
  R_0(y)=4y+b_1(b_1-2),\\
  R_{-1}(y)&=-i(a_1+a_2-a_3-a_4)y-i(b_1-2)(a_1a_2-a_3a_4),\\
  b_1&\eqdef\sum_{j=1}^4a_j,\quad
  (a_3,a_4)\eqdef(a_1^*,a_2^*)\ \text{ or }\ (a_2^*,a_1^*).
\end{align}
These can be rewritten as
\begin{align}
  \mathcal{E}_n(\bm{\lambda})&=n(n+2\text{Re}(a_1+a_2)-1),\\
  R_0(y)&=4y
  +4\text{Re}(a_1+a_2)\bigl(\text{Re}(a_1+a_2)-1\bigr),\\
  R_{-1}(y)&=2\text{Im}(a_1+a_2)y
  +4\bigl(\text{Re}(a_1+a_2)-1\bigr)\text{Im}(a_1a_2).
\end{align}

\paragraph{eigenfunctions}
\begin{align}
  \phi_0(x\,;\bm{\lambda})&\eqdef
  |\Gamma(a_1+ix)\Gamma(a_2+ix)|
  =\sqrt{\Gamma(a_1+ix)\Gamma(a_2+ix)\Gamma(a_3-ix)\Gamma(a_4-ix)},
  \label{Hahnphi0}\\
  P_n(\eta\,;\bm{\lambda})&=p_n(x\,;a_1,a_2,a_3,a_4)\n
  &\eqdef
  i^n\frac{(a_1+a_3)_n(a_1+a_4)_n}{n!}\,
  {}_3F_2\Bigl(\genfrac{}{}{0pt}{}{-n,\,n+a_1+a_2+a_3+a_4-1,\,a_1+ix}
  {a_1+a_3,\,a_1+a_4}\Bigm|1\Bigr),
  \label{defcH}
\end{align}
which are symmetric under $a_1\leftrightarrow a_2$ and
$a_3\leftrightarrow a_4$ separately.
\begin{align}
  c_n&=\frac{(n+b_1-1)_n}{n!},\\
  a_n^{\text{rec}}&=i\biggl(a_1
  -\frac{(n+b_1-1)(n+a_1+a_3)(n+a_1+a_4)}{(2n+b_1-1)(2n+b_1)}\n
  &\qquad\qquad
  +\frac{n(n+a_2+a_3-1)(n+a_2+a_4-1)}{(2n+b_1-2)(2n+b_1-1)}\biggr),\\
  b_n^{\text{rec}}&=\frac{n(n+b_1-2)\prod_{j=1}^2\prod_{k=3}^4(n+a_j+a_k-1)}
  {(2n+b_1-3)(2n+b_1-2)^2(2n+b_1-1)},\\
  f_n(\bm{\lambda})&=n+b_1-1,\quad
  b_n(\bm{\lambda})=n+1.
\end{align}

\paragraph{annihilation/creation operators and commutation relations}
\begin{align}
  \alpha_{\pm}(\mathcal{H})&=1\pm 2\sqrt{\mathcal{H}'},\quad
  \mathcal{H}'\eqdef\mathcal{H}+\tfrac14(b_1-1)^2,\\
  \mathcal{N}&=\sqrt{\mathcal{H}'}-\tfrac12(b_1-1)\quad(\text{for }b_1>1),\\
  [\mathcal{H},a^{(\pm)}]&=a^{(\pm)}(1\pm 2\sqrt{\mathcal{H}'}).
\end{align}
The annihilation/creation operators \eqref{a^{(pm)}} and their
commutation relation \eqref{[a-,a+]} are not so simplified because
$b^{\text{rec}}_{n+1}-b^{\text{rec}}_{n}
=(\text{quartic polynomial in $n$})/(\text{cubic polynomial in $n$})$
has a lengthy expression.

\paragraph{coherent state}
\begin{equation}
  \psi(\alpha,x\,;\bm{\lambda})=\phi_0(x\,;\bm{\lambda})
  \sum_{n=0}^{\infty}
  \frac{(b_1)_{2n}\,\alpha^n}{\prod_{j=1}^2\prod_{k=3}^4(a_j+a_k)_n}\,
  P_n(\eta(x)\,;\bm{\lambda}).
  \label{31coh}
\end{equation}
The r.h.s is symmetric under $a_1\leftrightarrow a_2$ and
$a_3\leftrightarrow a_4$ separately.
We are not aware if a concise summation formula exists or not.
Several non-symmetric generating functions for the continuous Hahn
polynomial are given in \cite{koeswart}.

\paragraph{orthogonality}
\begin{gather}
  \int_{-\infty}^{\infty}\phi_0(x\,;\bm{\lambda})^2
  P_n(\eta\,;\bm{\lambda})P_m(\eta\,;\bm{\lambda})dx
  =2\pi\,
  \frac{\prod_{j=1}^2\prod_{k=3}^4\Gamma(n+a_j+a_k)}
  {n!\,(2n+b_1-1)\Gamma(n+b_1-1)}\,\delta_{nm},\\
  \frac{1}{h_0(\bm{\lambda})}=
  \frac{\Gamma(b_1)}{2\pi\prod_{j=1}^2\prod_{k=3}^4\Gamma(a_j+a_k)}\,,\quad
  \frac{h_0(\bm{\lambda})}{h_n(\bm{\lambda})}=
  \frac{b_1+2n-1}{b_1+n-1}
  \frac{n!\,(b_1)_n}{\prod_{j=1}^2\prod_{k=3}^4(a_j+a_k)_n}\,.
\end{gather}

\subsection{Meixner-Pollaczek [KS1.7]}
\label{[KS1.7]}

In previous works \cite{os4,os7,degruij}, the parameter $\phi$ was
fixed to $\pi/2$.
Here we treat the most general case $0<\phi<\pi$.

\paragraph{parameters and potential function}
\begin{equation}
  \bm{\lambda}\eqdef a,\quad \bm{\delta}=\tfrac12,\quad
  \phi\ \ (0<\phi<\pi);\quad a>0; \quad
  V(x\,;\bm{\lambda})\eqdef e^{i(\frac{\pi}{2}-\phi)}(a+ix).
\end{equation}

\paragraph{shape invariance and closure relation}
\begin{align}
  \mathcal{E}_n(\bm{\lambda})&=2n\sin\phi,\\
  R_1(y)&=0,\quad
  R_0(y)=4\sin^2\phi,\quad
  R_{-1}(y)=2y\cos\phi+2a\sin2\phi.
\end{align}

\paragraph{eigenfunctions}
\begin{align}
  \phi_0(x\,;\bm{\lambda})&\eqdef
  e^{(\phi-\frac{\pi}{2})x}|\Gamma(a+ix)|,
  \label{Meixphi0}\\
  P_n(\eta\,;\bm{\lambda})&=P_n^{(a)}(x\,;\phi)
  \eqdef\frac{(2a)_n}{n!}\,e^{in\phi}
  {}_2F_1\Bigl(\genfrac{}{}{0pt}{}{-n,\,a+ix}{2a}\Bigm|
  1-e^{-2i\phi}\Bigr),
  \label{defMP}\\
  c_n&=\frac{(2\sin\phi)^n}{n!}\quad
  a_n^{\text{rec}}=-\frac{n+a}{\tan\phi},\quad
  b_n^{\text{rec}}=\frac{n(n+2a-1)}{(2\sin\phi)^2},
  \label{recKS1.7}\\
  f_n(\bm{\lambda})&=2\sin\phi,\quad
  b_n(\bm{\lambda})=n+1.
\end{align}
The polynomial has the following symmetry
$P_n^{(a)}(x\,;-\phi)=P_n^{(a)}(-x\,;\phi)$.

\paragraph{annihilation/creation operators and commutation relations}
\begin{align}
  \alpha_{\pm}(\mathcal{H})&=\pm 2\sin\phi,\quad
  \mathcal{N}=\frac{1}{2\sin\phi}\,\mathcal{H},\\
  a^{(\pm)}&=\frac{\pm 1}{4\sin\phi}[\mathcal{H},\eta]+\frac12\eta
  +\frac{\cos\phi}{4\sin^2\phi}(\mathcal{H}+2a\sin\phi),\\
  b^{\text{rec}}_{n+1}-b^{\text{rec}}_{n}&=\frac{n+a}{2\sin^2\phi},\\
  [\mathcal{H},a^{(\pm)}]&=\pm 2\sin\phi\,\,a^{(\pm)},\\
  [a^{(-)},a^{(+)}]&=\frac{1}{4\sin^3\phi}(\mathcal{H}+2a\sin\phi).
\end{align}
\begin{align}
  \mathfrak{su}(1,1)\ \text{algebra}:\quad& J^{\pm}=2\sin\phi\,\,a^{(\pm)},
  \quad J^3=\frac{1}{2\sin\phi}(\mathcal{H}+2a\sin\phi),\n
  &[J^3,J^{\pm}]=\pm J^{\pm},\quad [J^-,J^+]=2J^3.
\end{align}
The $\mathfrak{su}(1,1)$ or $\mathfrak{sl}(2,\mathbb{R})$ algebra
reported before \cite{os7,degruij} is a special case of the present one.

\paragraph{$\bm{\lambda}$-shift operators}
For the special case of $\phi=\pi/2$ the annihilation/creation
operators are closely related to the $\mathcal{A}$ and
$\mathcal{A}^\dagger$ operators:
\begin{align}
  &a^{(+)}=\mathcal{A}^{\dagger}X,\quad
  X=\tfrac14(S_++S_-),\\
  &a^{(-)}=X^{\dagger}\mathcal{A},\quad
  X^{\dagger}=\tfrac14(S_+^{\dagger}+S_-^{\dagger}),
\end{align}
\begin{align}
  \phi_0(x\,;\bm{\lambda}+\bm{\delta})^{-1}\,X(\bm{\lambda})\,\,
  \phi_0(x\,;\bm{\lambda})\cdot P_n(\eta\,;\bm{\lambda})
  &=\tfrac12P_n(\eta\,;\bm{\lambda}+\bm{\delta}),\\
  \phi_0(x\,;\bm{\lambda})^{-1}\,X(\bm{\lambda})^{\dagger}\,\,
  \phi_0(x\,;\bm{\lambda}+\bm{\delta})\cdot
  P_n(\eta\,;\bm{\lambda}+\bm{\delta})
  &=\tfrac14(n+2a)P_n(\eta\,;\bm{\lambda}).
\end{align}

\paragraph{coherent state}
The coherent state gives a simple generating function, which generalises
the previous result \cite{os7}:
\begin{align}
  \psi(\alpha,x\,;\bm{\lambda})&=\phi_0(x\,;\bm{\lambda})
  \sum_{n=0}^{\infty}\frac{(2\sin\phi)^n\alpha^n}{(2a)_n}\,
  P_n(\eta(x)\,;\bm{\lambda})\n
  &=\phi_0(x\,;\bm{\lambda})\,e^{ i\alpha(1-e^{2i\phi})}
  {}_1F_1\Bigl(\genfrac{}{}{0pt}{}{a+ix}{2a}\Bigm|-4i\alpha\sin^2\phi
  \Bigr).
  \label{32coh}
\end{align}

\paragraph{orthogonality}
\begin{gather}
  \int_{-\infty}^{\infty}\phi_0(x\,;\bm{\lambda})^2
  P_n(\eta\,;\bm{\lambda})P_m(\eta\,;\bm{\lambda})dx
  =2\pi\,\frac{\Gamma(n+2a)}{n!\,(2\sin\phi)^{2a}}\,\delta_{nm},\\
  \frac{1}{h_0(\bm{\lambda})}=
  \frac{(2\sin\phi)^{2a}}{2\pi\Gamma(2a)}\,,\quad
  \frac{h_0(\bm{\lambda})}{h_n(\bm{\lambda})}=
  \frac{n!}{(2a)_n}\,.
\end{gather}

The exact solvability of the continuous Hahn and Meixner-Pollaczek
polynomials for the full parameters are discussed in \cite{newqes}
in connection with their further deformation to give another example
of quasi exactly solvable system.

\section{$\eta(x)=x^2$}
\label{secx2}
\setcounter{equation}{0}

In all the examples in this section, we have
\begin{equation}
  \eta(x)=x^2,\quad 0<x<\infty,\quad \gamma=1,\quad
  \kappa=1, \quad \varphi(x)=2x.
  \label{first4}
\end{equation}

\subsection{Wilson [KS1.1]}
\label{[KS1.1]}

The Wilson polynomial is the most general one in this category.
The parameters $a_1$,\ldots, $a_4$ were restricted to real positive
values in previous works \cite{os4,os5,os6,os7}.
The generic situation to be discussed in this paper is
\begin{equation}
  \{a_1^*,a_2^*,a_3^*,a_4^*\}=\{a_1,a_2,a_3,a_4\} \quad (\text{as a set}),
  \quad \text{Re}\,a_i>0\ \ (1\leq i\leq 4).
  \label{condKS1.1}
\end{equation}

\paragraph{parameters and potential function}
\begin{equation}
  \bm{\lambda}\eqdef(a_1,a_2,a_3,a_4),
  \ \bm{\delta}=(\tfrac12,\tfrac12,\tfrac12,\tfrac12);
  \ \ V(x\,;\bm{\lambda})\eqdef
  \frac{(a_1+ix)(a_2+ix)(a_3+ix)(a_4+ix)}{2ix(2ix+1)}.
\end{equation}

\paragraph{shape invariance and closure relation}
\begin{align}
  \mathcal{E}_n(\bm{\lambda})&=n(n+b_1-1),\\
  R_1(y)&=2,\quad
  R_0(y)=4y+b_1(b_1-2),\quad
  R_{-1}(y)=-2y^2+(b_1-2b_2)y+(2-b_1)b_3,\\
  b_1&\eqdef\sum_{j=1}^4a_j,\quad
  b_2\eqdef\!\!\sum_{1\leq j<k\leq 4}a_ja_k,\quad
  b_3\eqdef\!\!\!\sum_{1\leq j<k<l\leq 4}a_ja_ka_l.
\end{align}

\paragraph{eigenfunctions}
\begin{align}
  \phi_0(x\,;\bm{\lambda})&\eqdef
  \biggl|\frac{\prod_{j=1}^4\Gamma(a_j+ix)}{\Gamma(2ix)}\biggr|,
  \label{Wilsonphi0}\\[4pt]
  P_n(\eta\,;\bm{\lambda})&=W_n(x^2\,;a_1,a_2,a_3,a_4)\n
  &\eqdef(a_1+a_2)_n(a_1+a_3)_n(a_1+a_4)_n\n
  &\qquad\qquad\times
  {}_4F_3\Bigl(
  \genfrac{}{}{0pt}{}{-n,\,n+\sum_{j=1}^4a_j-1,\,a_1+ix,\,a_1-ix}
  {a_1+a_2,\,a_1+a_3,\,a_1+a_4}\Bigm|1\Bigr),
  \label{defW}
\end{align}
which are symmetric under the permutations of $(a_1,a_2,a_3,a_4)$.
\begin{align}
  c_n&=(-1)^n(n+b_1-1)_n,\\
  a_n^{\text{rec}}&=
  \frac{(n+b_1-1)\prod_{j=2}^4(n+a_1+a_j)}{(2n+b_1-1)(2n+b_1)}
  +\frac{n\prod_{2\leq j<k\leq 4}(n+a_j+a_k-1)}{(2n+b_1-2)(2n+b_1-1)}
  -a_1^2,\\
  b_n^{\text{rec}}&=\frac{n(n+b_1-2)\prod_{1\leq j<k\leq 4}(n+a_j+a_k-1)}
  {(2n+b_1-3)(2n+b_1-2)^2(2n+b_1-1)},\\
  f_n(\bm{\lambda})&=-n(n+b_1-1),\quad
  b_n(\bm{\lambda})=-1.
\end{align}

\paragraph{annihilation/creation operators and commutation relations}
\begin{align}
  \alpha_{\pm}(\mathcal{H})&=1\pm 2\sqrt{\mathcal{H}'},\quad
  \mathcal{H}'=\mathcal{H}+\tfrac14(b_1-1)^2,\\
  \mathcal{N}&=\sqrt{\mathcal{H}'}-\tfrac12(b_1-1)\quad(\text{for }b_1>1),\\
  [\mathcal{H},a^{(\pm)}]&=a^{(\pm)}(1\pm 2\sqrt{\mathcal{H}'}).
\end{align}
The annihilation/creation operators \eqref{a^{(pm)}} and their
commutation relation \eqref{[a-,a+]} are not so simplified because
the expression
$b^{\text{rec}}_{n+1}-b^{\text{rec}}_{n}
=(\text{10-th degree polynomial in $n$})/
(\text{7-th degree}$ $\text{polynomial in $n$})$
is quite complicated.

\paragraph{coherent state}
\begin{equation}
  \psi(\alpha,x\,;\bm{\lambda})=\phi_0(x\,;\bm{\lambda})
  \sum_{n=0}^{\infty}\frac{(-1)^n(b_1)_{2n}\,\alpha^n}
  {n!\prod_{1\leq j<k\leq 4}(a_j+a_k)_n}\,
  P_n(\eta(x)\,;\bm{\lambda}).
  \label{41coh}
\end{equation}
The r.h.s. is symmetric under the permutations of $(a_1,a_2,a_3,a_4)$.
It is not known to us if a concise summation formula exists or not.
Several non-symmetric generating functions for the Wilson polynomial
are given in \cite{koeswart}.

\paragraph{orthogonality}
\begin{gather}
  \int_0^{\infty}\phi_0(x\,;\bm{\lambda})^2
  P_n(\eta\,;\bm{\lambda})P_m(\eta\,;\bm{\lambda})dx
  =2\pi n!\,(n+b_1-1)_n\,
  \frac{\prod_{1\leq j<k\leq 4}\Gamma(n+a_j+a_k)}{\Gamma(2n+b_1)}
  \,\delta_{nm},\\
  \frac{1}{h_0(\bm{\lambda})}=
  \frac{\Gamma(b_1)}{2\pi\prod_{1\leq j<k\leq 4}\Gamma(a_j+a_k)}\,,\quad
  \frac{h_0(\bm{\lambda})}{h_n(\bm{\lambda})}=
  \frac{b_1+2n-1}{b_1+n-1}
  \frac{(b_1)_n}{n!\prod_{1\leq j<k\leq 4}(a_j+a_k)_n}\,.
\end{gather}

\subsection{continuous dual Hahn [KS1.3]}
\label{[KS1.3]}

This is a restricted case of the Wilson polynomial with $a_4=0$.
In previous works \cite{os4,os5,os6,os7}, the parameters $a_1$, $a_2$
and $a_3$ were real and positive. Now they are
$\{a_1^*,a_2^*,a_3^*\}=\{a_1,a_2,a_3\}$, as a set and Re\,$a_i>0$.
This is dual to the continuous Hahn \S\ref{[KS1.4]} in the sense that
the roles of $\eta(x)$ and $\mathcal{E}_n$ are interchanged.
For the continuous Hahn, $\eta(x)=x$ and $\mathcal{E}_n$ is quadratic
in $n$, whereas $\eta(x)$ is quadratic in $x$ and $\mathcal{E}_n=n$
for the dual Hahn.
The duality has sharper meaning for polynomials with discrete
orthogonality measures, see for example \cite{os12}.

\paragraph{parameters and potential function}
\begin{equation}
  \bm{\lambda}\eqdef(a_1,a_2,a_3),\quad
  \bm{\delta}=(\tfrac12,\tfrac12,\tfrac12);\quad
  V(x\,;\bm{\lambda})\eqdef\frac{(a_1+ix)(a_2+ix)(a_3+ix)}{2ix(2ix+1)}.
\end{equation}

\paragraph{shape invariance and closure relation}
\begin{align}
  \mathcal{E}_n(\bm{\lambda})&=n,\\
  R_1(y)&=0,\quad
  R_0(y)=1,\quad
  R_{-1}(y)=-2y^2+(1-2b_1)y-b_2,\\
  b_1&\eqdef a_1+a_2+a_3,\quad b_2\eqdef a_1a_2+a_1a_3+a_2a_3.
\end{align}

\paragraph{eigenfunctions}
\begin{align}
  \phi_0(x\,;\bm{\lambda})&\eqdef
  \biggl|\frac{\prod_{j=1}^3\Gamma(a_j+ix)}{\Gamma(2ix)}\biggr|,
  \label{dualHahnphi0}\\[4pt]
  P_n(\eta\,;\bm{\lambda})&=S_n(x^2\,;a_1,a_2,a_3)\n
  &\eqdef(a_1+a_2)_n(a_1+a_3)_n
  \ {}_3F_2\Bigl(\genfrac{}{}{0pt}{}{-n,\,a_1+ix,\,a_1-ix}
  {a_1+a_2,\,a_1+a_3}\Bigm|1\Bigr),
  \label{defcdH}
\end{align}
which are symmetric under the permutations of $(a_1,a_2,a_3)$.
\begin{align}
  c_n&=(-1)^n,\\
  a_n^{\text{rec}}&=(n+a_1+a_2)(n+a_1+a_3)+n(n+a_2+a_3-1)-a_1^2,\\
  b_n^{\text{rec}}&=n\prod_{1\leq j<k\leq 3}(n+a_j+a_k-1),\\
  f_n(\bm{\lambda})&=-n,\quad
  b_n(\bm{\lambda})=-1.
\end{align}

\paragraph{annihilation/creation operators and commutation relations}
\begin{align}
  \alpha_{\pm}(\mathcal{H})&=\pm 1,\qquad \mathcal{N}=\mathcal{H},\\
  a^{(\pm)}&=\pm\tfrac12[\mathcal{H},\eta]+\tfrac12\eta
  -\mathcal{H}^2-(b_1-\tfrac12)\mathcal{H}-\tfrac12b_2,\\
  b^{\text{rec}}_{n+1}-b^{\text{rec}}_{n}
  &=4n^3+3(2b_1-1)n^2+\bigl(2b_1(b_1-1)+2b_2+1\bigr)n+b_1b_2-a_1a_2a_3.
\end{align}
The interesting algebra, reported in \cite{os7}, with $\mathcal{H}^3$
non-linearity on the r.h.s. of \eqref{cdHahnalge} is valid for the full
parameter range:
\begin{align}
  [\mathcal{H},a^{(\pm)}]&=\pm a^{(\pm)},\\
  [a^{(-)},a^{(+)}]&=
  4\mathcal{H}^3+3(2b_1-1)\mathcal{H}^2
  +\bigl(2b_1(b_1-1)+2b_2+1\bigr)\mathcal{H}+b_1b_2-a_1a_2a_3.
  \label{cdHahnalge}
\end{align}

\paragraph{$\bm{\lambda}$-shift operators}
\begin{align}
  X&=-iS_+T_+ +\Bigl(x-iV(x-\tfrac{i}{2})^*
  -i\,\frac{\prod_{j=1}^3(2a_j-1)}{8(1+x^2)}\Bigr)S_+ \n
  &\quad +iS_-T_- +\Bigl(x+iV(x-\tfrac{i}{2})\,\,
  +i\,\frac{\prod_{j=1}^3(2a_j-1)}{8(1+x^2)}\Bigr)S_-,
\end{align}
\begin{align}
  \phi_0(x\,;\bm{\lambda}+\bm{\delta})^{-1}\,X(\bm{\lambda})\,\,
  \phi_0(x\,;\bm{\lambda})\cdot P_n(\eta\,;\bm{\lambda})
  &=P_n(\eta\,;\bm{\lambda}+\bm{\delta}),\\
  \phi_0(x\,;\bm{\lambda})^{-1}\,X(\bm{\lambda})^{\dagger}\,\,
  \phi_0(x\,;\bm{\lambda}+\bm{\delta})\cdot
  P_n(\eta\,;\bm{\lambda}+\bm{\delta})
  &=\!\!\prod_{1\leq j<k\leq 3}(n+a_j+a_k)\cdot P_n(\eta\,;\bm{\lambda}).
\end{align}

\paragraph{coherent state}
\begin{equation}
  \psi(\alpha,x\,;\bm{\lambda})=\phi_0(x\,;\bm{\lambda})
  \sum_{n=0}^{\infty}\frac{(-1)^n\alpha^n}
  {n!\prod_{1\leq j<k\leq 3}(a_j+a_k)_n}\,
  P_n(\eta(x)\,;\bm{\lambda}).
  \label{42coh}
\end{equation}
The r.h.s. is symmetric under the permutations of $(a_1,a_2,a_3)$.
We are not aware if a concise summation formula exists or not.
Several non-symmetric generating functions for the continuous dual
Hahn polynomial are given in \cite{koeswart}.

\paragraph{orthogonality}
\begin{gather}
  \int_0^{\infty}\phi_0(x\,;\bm{\lambda})^2
  P_n(\eta\,;\bm{\lambda})P_m(\eta\,;\bm{\lambda})dx
  =2\pi n!\,\prod_{1\leq j<k\leq 3}\Gamma(n+a_j+a_k)\cdot\delta_{nm},\\
  \frac{1}{h_0(\bm{\lambda})}=
  \frac{1}{2\pi\prod_{1\leq j<k\leq 3}\Gamma(a_j+a_k)}\,,\quad
  \frac{h_0(\bm{\lambda})}{h_n(\bm{\lambda})}=
  \frac{1}{n!\prod_{1\leq j<k\leq 3}(a_j+a_k)_n}\,.
\end{gather}

\section{$\eta(x)=\cos x$}
\label{seccosx}
\setcounter{equation}{0}

In all the examples in this section, we have\footnote{
We have changed the sign of $\varphi(x)$ from \cite{os7}.}
\begin{equation}
  \eta(x)=\cos x,\quad 0<x<\pi,\quad \gamma=\log q,\quad
  \kappa=q^{-1}, \quad \varphi(x)=2\sin x.
  \label{first5}
\end{equation}
Throughout this paper $q$ is always in the range $0<q<1$ and
this will not be indicated.
It is convenient to introduce a complex variable $z=e^{ix}$.
Then the shift operator $e^{\gamma p}$ can be written as
\begin{equation}
  e^{\gamma p}=e^{-i\gamma\frac{d}{dx}}=q^{z\frac{d}{dz}},
\end{equation}
whose action on a function of $x$ can be expressed as $z\to qz$:
\[
  e^{\gamma p}f(x)=f(x-i\gamma)=q^{z\frac{d}{dz}}\check{f}(z)=\check{f}(qz),
  \quad \text{with}\  f(x)=\check{f}(z).
\]
Note that $\gamma<0$.

\subsection{Askey-Wilson [KS3.1]}
\label{[KS3.1]}

The Askey-Wilson polynomial is the most general one with the maximal
number of parameters, four.
All the other polynomials in this  section are obtained
by restricting the parameters $a_1$,\ldots,$a_4$, in one way or another.
In previous publications \cite{os4,os5,os6,os7} these restricted
polynomials were not discussed individually, since their exact
solvability is a simple corollary of that of the Askey-Wilson.
However, the simpler structure of the restricted ones would give rise
to simple energy spectrum and interesting and tractable forms of the
dynamical symmetry algebras and coherent states, etc., as exemplified
by the continuous $q$-Hermite polynomial \S\ref{[KS3.26]}, which has
$a_1=a_2=a_3=a_4=0$. It gives a most natural realisation of the
$q$-oscillator algebra \cite{os11}.

\paragraph{parameters and potential function}
\begin{gather}
  q^{\bm{\lambda}}\eqdef(a_1,a_2,a_3,a_4),\quad
  \bm{\delta}=(\tfrac12,\tfrac12,\tfrac12,\tfrac12),\quad q;\\[4pt]
  V(x\,;\bm{\lambda})\eqdef\frac{(1-a_1z)(1-a_2z)(1-a_3z)(1-a_4z)}
  {(1-z^2)(1-qz^2)},\quad z=e^{ix}.
\end{gather}
The parameters have to satisfy the conditions
\begin{equation}
  \{a_1^*,a_2^*,a_3^*,a_4^*\}=\{a_1,a_2,a_3,a_4\} \quad (\text{as a set}),
  \qquad |a_i|<1,\quad i=1,\ldots,4.
  \label{condKS3.1}
\end{equation}
In previous works \cite{os4,os5,os6,os7} only the real parameters
$a_i\in\mathbb{R}$ were discussed.

\paragraph{shape invariance and closure relation}
\begin{align}
  \mathcal{E}_n(\bm{\lambda})&=(q^{-n}-1)(1-b_4q^{n-1}),\\
  R_1(y)&=(q^{-\frac12}-q^{\frac12})^2y',\quad y'\eqdef y+1+q^{-1}b_4,\\
  R_0(y)&=(q^{-\frac12}-q^{\frac12})^2
  \bigl(y^{\prime\,2}-(1+q^{-1})^2b_4\bigr),\\
  R_{-1}(y)&=-\tfrac12(q^{-\frac12}-q^{\frac12})^2
  \bigl((b_1+q^{-1}b_3)y'-(1+q^{-1})(b_3+q^{-1}b_1b_4)\bigr),\\
  b_1&\eqdef\sum_{j=1}^4a_j,\quad
  b_3\eqdef\!\!\!\sum_{1\leq j<k<l\leq 4}a_ja_ka_l,\quad
  b_4\eqdef a_1a_2a_3a_4.
\end{align}

\paragraph{eigenfunctions}
\begin{align}
  \phi_0(x\,;\bm{\lambda})&\eqdef
  \biggl|\frac{(e^{2ix}\,;q)_{\infty}}
  {\prod_{j=1}^4(a_je^{ix}\,;q)_{\infty}}\biggr|,
  \label{AWilsonphi0}\\[4pt]
  P_n(\eta\,;\bm{\lambda})&=p_n(\cos x\,;a_1,a_2,a_3,a_4|q)\n
  &\eqdef a_1^{-n}(a_1a_2,a_1a_3,a_1a_4\,;q)_n
  \times
  {}_4\phi_3\Bigl(\genfrac{}{}{0pt}{}{q^{-n},\,a_1a_2a_3a_4q^{n-1},\,
  a_1e^{ix},\,a_1e^{-ix}}{a_1a_2,\,a_1a_3,\,a_1a_4}\Bigm|q\,;q\Bigr),
  \label{defAW}
\end{align}
which are symmetric under the permutations of $(a_1,a_2,a_3,a_4)$.
\begin{align}
  c_n&=2^n(b_4q^{n-1}\,;q)_n,\\
  a_n^{\text{rec}}&=\frac12\biggl(a_1+a_1^{-1}
  -\frac{(1-b_4q^{n-1})\prod_{j=2}^4(1-a_1a_jq^n)}
  {a_1(1-b_4q^{2n-1})(1-b_4q^{2n})}\n
  &\qquad\qquad\qquad\quad
  -\frac{a_1(1-q^n)\prod_{2\leq j<k\leq 4}(1-a_ja_kq^{n-1})}
  {(1-b_4q^{2n-2})(1-b_4q^{2n-1})}\biggr),\\
  b_n^{\text{rec}}&=\frac{(1-q^n)(1-b_4q^{n-2})
  \prod_{1\leq j<k\leq 4}(1-a_ja_kq^{n-1})}
  {4(1-b_4q^{2n-3})(1-b_4q^{2n-2})^2(1-b_4q^{2n-1})},\\
  f_n(\bm{\lambda})&=q^{\frac{n}{2}}(q^{-n}-1)(1-b_4q^{n-1}),\quad
  b_n(\bm{\lambda})=q^{-\frac{n+1}{2}}.
\end{align}

\paragraph{annihilation/creation operators and commutation relations}
\begin{align}
  \alpha_{\pm}(\mathcal{H})
  &=\tfrac12(q^{-\frac12}-q^{\frac12})^2\mathcal{H}'
  \pm\tfrac12(q^{-1}-q)\sqrt{\mathcal{H}^{\prime\,2}-4q^{-1}b_4}\,,\quad
  \mathcal{H}'=\mathcal{H}+1+q^{-1}b_4\,,\\
  q^{\mathcal{N}}&=\frac{q}{2b_4}\left(\mathcal{H}'
  -\sqrt{\mathcal{H}^{\prime\,2}-4q^{-1}b_4}\,\right)\quad
  (\text{for }0<b_4<q),\\
  [\mathcal{H},a^{(\pm)}]&=\tfrac12a^{(\pm)}\Bigl(
  (q^{-\frac12}-q^{\frac12})^2\mathcal{H}'
  \pm(q^{-1}-q)\sqrt{\mathcal{H}^{\prime\,2}-4q^{-1}b_4}\Bigr).
\end{align}
The annihilation/creation operators \eqref{a^{(pm)}} and their
commutation relation \eqref{[a-,a+]} are not simplified at all.
The expression
$b^{\text{rec}}_{n+1}-b^{\text{rec}}_{n}
=q^n\times(\text{12-th degree polynomial in $q^n$})/
(\text{6-th degree}$ $\text{polynomial in $q^{2n}$})$
is very complicated.

\paragraph{coherent state}
\begin{equation}
  \psi(\alpha,x\,;\bm{\lambda})=\phi_0(x\,;\bm{\lambda})
  \sum_{n=0}^{\infty}\frac{2^n(b_4\,;q)_{2n}\alpha^n}
  {(q\,;q)_n\prod_{1\leq j<k\leq 4}(a_ja_k\,;q)_n}\,
  P_n(\eta(x)\,;\bm{\lambda}).
  \label{51coh}
\end{equation}
The r.h.s. is symmetric under the permutations of $(a_1,a_2,a_3,a_4)$.
We are not aware if a concise summation formula exists or not.
Several non-symmetric generating functions for the Askey-Wilson
polynomial are given in \cite{koeswart}.

\paragraph{orthogonality}
\begin{gather}
  \int_0^{\pi}\phi_0(x\,;\bm{\lambda})^2
  P_n(\eta\,;\bm{\lambda})P_m(\eta\,;\bm{\lambda})dx
  =2\pi\,
  \frac{(b_4q^{n-1}\,;q)_n(b_4q^{2n}\,;q)_{\infty}}
  {(q^{n+1}\,;q)_{\infty}\prod_{1\leq j<k\leq 4}(a_ja_kq^n\,;q)_{\infty}}
  \,\delta_{nm},\\
  \frac{1}{h_0(\bm{\lambda})}=
  \frac{(q\,;q)_{\infty}\prod_{1\leq j<k\leq 4}(a_ja_k\,;q)_{\infty}}
  {2\pi(b_4\,;q)_{\infty}}\,,\quad
  \frac{h_0(\bm{\lambda})}{h_n(\bm{\lambda})}=
  \frac{1-b_4q^{2n-1}}{1-b_4q^{n-1}}
  \frac{(b_4\,;q)_n}{(q\,;q)_n\prod_{1\leq j<k\leq 4}(a_ja_k\,;q)_n}\,.
\end{gather}

\subsubsection{Askey-Wilson $\to$ Wilson}

The Wilson polynomial is obtained from the Askey-Wilson polynomial by
a $q\uparrow1$ limit. Here we present a dictionary of the correspondence
for future reference.
Let us first introduce a new coordinate $x'$ for the Wilson polynomial
as the rescaled one of the variable $x$ ($0<x<\pi$) of the Askey-Wilson
polynomial:
\begin{equation}
  x'=\frac{L}{\pi}x,
  \ \ \bigl(\Rightarrow 0<x'<L,\ \ p'=\frac{\pi}{L}p\bigr),\quad
  \gamma=-\frac{\pi}{L},\quad
  \bm{\lambda}=(a'_1,a'_2,a'_3,a'_4),
\end{equation}
in which $L$ is related to $q$ as $q=e^{-\pi/L}$.
This entails
\begin{equation}
  e^{\gamma p}=e^{-p'}
\end{equation}
and the following limit formulas as $L\to\infty$ or $q\to 1$:
(The superscript {\scriptsize W} denote the corresponding quantity
for the Wilson polynomial.)
\begin{align}
  &\lim_{L\to\infty}\frac{V(x\,;\bm{\lambda})}{(1-q)^2}
  =V^{\text{W}}(x'\,;\bm{\lambda})^*,\\
  &\lim_{L\to\infty}\frac{\mathcal{H}(\bm{\lambda})}{(1-q)^2}
  =\mathcal{H}^{\text{W}}(\bm{\lambda}),\quad
  \lim_{L\to\infty}\frac{\mathcal{E}_n(\bm{\lambda})}{(1-q)^2}
  =\mathcal{E}_n^{\text{W}}(\bm{\lambda}),\\
  &\lim_{L\to\infty}(q\,;q)_{\infty}^3(1-q)^{3-\sum_ja'_j}
  \phi_0(x\,;\bm{\lambda})
  =\phi_0^{\text{W}}(x'\,;\bm{\lambda}),\quad
  \lim_{L\to\infty}\frac{\varphi(x)}{1-q}
  =\varphi^{\text{W}}(x'),\\
  &\lim_{L\to\infty}\frac{P_n(\eta(x)\,;\bm{\lambda})}{(1-q)^{3n}}
  =P_n^{\text{W}}(\eta^{\text{W}}(x')\,;\bm{\lambda}),\\
  &\lim_{L\to\infty}(1-q)\mathcal{F}(\bm{\lambda})
  =-\mathcal{F}^{\text{W}}(\bm{\lambda}),\quad
  \lim_{L\to\infty}\frac{f_n(\bm{\lambda})}{(1-q)^2}
  =-f_n^{\text{W}}(\bm{\lambda}),\\
  &\lim_{L\to\infty}\frac{\mathcal{B}(\bm{\lambda})}{(1-q)^3}
  =-\mathcal{B}^{\text{W}}(\bm{\lambda}),\quad
  \lim_{L\to\infty}b_n(\bm{\lambda})
  =-b_n^{\text{W}}(\bm{\lambda}).
\end{align}

\subsection{continuous dual $q$-Hahn [KS3.3]}
\label{[KS3.3]}

The continuous dual $q$-Hahn polynomial is obtained by restricting
$a_4=0$ in the Askey-Wilson polynomial \S\ref{[KS3.1]}.
This restriction renders
the energy spectrum to a simple form $\mathcal{E}_n=q^{-n}-1$ for all
the restricted polynomials in section \ref{seccosx} 
except for the continuous $q$-Jacobi polynomial \S\ref{[KS3.10]} and the
continuous $q$-Hahn  polynomial \S\ref{[KS3.4]}.
For these the commutation relations of $\mathcal{H}$ and of $a^{(\pm)}$
is the same \eqref{KS3.3[H,apm]}, \eqref{KS3.8[H,apm]},
\eqref{KS3.18[H,apm]}, \eqref{KS3.26[H,apm]} and \eqref{KS3.19[H,apm]}.
They can be expressed as $q$-deformed commutators \eqref{KS3.3qc},
\eqref{KS3.8qc}, \eqref{KS3.18qc}, \eqref{KS3.26qc} and \eqref{KS3.19qc}.
The commutation relation $[a^{(-)},a^{(+)}]$ or its deformation
becomes drastically simpler, as the number of parameters decreases.

\paragraph{parameters and potential function}
\begin{gather}
  q^{\bm{\lambda}}\eqdef(a_1,a_2,a_3),\quad
  \bm{\delta}=(\tfrac12,\tfrac12,\tfrac12),\quad q; \\[4pt]
  V(x\,;\bm{\lambda})\eqdef\frac{(1-a_1z)(1-a_2z)(1-a_3z)}
  {(1-z^2)(1-qz^2)},\quad z=e^{ix}.
\end{gather}
The parameters have to satisfy the conditions
\begin{equation}
  \{a_1^*,a_2^*,a_3^*\}=\{a_1,a_2,a_3\} \quad (\text{as a set}),
  \qquad |a_i|<1,\quad i=1,2,3.
  \label{condKS3.3}
\end{equation}

\paragraph{shape invariance and closure relation}
\begin{align}
  \mathcal{E}_n(\bm{\lambda})&=q^{-n}-1,\\
  R_1(y)&=(q^{-\frac12}-q^{\frac12})^2y',\quad
  y'\eqdef y+1,\quad
  R_0(y)=(q^{-\frac12}-q^{\frac12})^2y^{\prime\,2},\\
  R_{-1}(y)&=-\tfrac12(q^{-\frac12}-q^{\frac12})^2
  \bigl((b_1+q^{-1}b_3)y'-(1+q^{-1})b_3\bigr),\\
  b_1&\eqdef a_1+a_2+a_3,\quad b_2\eqdef a_1a_2+a_1a_3+a_2a_3,\quad
  b_3\eqdef a_1a_2a_3.
\end{align}

\paragraph{eigenfunctions}
\begin{align}
  \phi_0(x\,;\bm{\lambda})&\eqdef
  \biggl|\frac{(e^{2ix}\,;q)_{\infty}}
  {\prod_{j=1}^3(a_je^{ix}\,;q)_{\infty}}\biggr|,
  \label{qHahnphi0}\\[4pt]
  P_n(\eta\,;\bm{\lambda})&=p_n(\cos x\,;a_1,a_2,a_3|q)\n
  &\eqdef
  a_1^{-n}(a_1a_2,a_1a_3\,;q)_n\,
  {}_3\phi_2\Bigl(\genfrac{}{}{0pt}{}{q^{-n},\,
  a_1e^{ix},\,a_1e^{-ix}}{a_1a_2,\,a_1a_3}\Bigm|q\,;q\Bigr),
  \label{defcdqH}
\end{align}
which are symmetric under the permutations of $(a_1,a_2,a_3)$.
\begin{align}
  c_n&=2^n,\\
  a_n^{\text{rec}}&=\tfrac12\bigl(a_1+a_1^{-1}
  -a_1^{-1}(1-a_1a_2q^n)(1-a_1a_3q^n)
  -a_1(1-q^n)(1-a_2a_3q^{n-1})\bigr),\\
  b_n^{\text{rec}}&=\tfrac14(1-q^n)\prod_{1\leq j<k\leq 3}(1-a_ja_kq^{n-1}),\\
  f_n(\bm{\lambda})&=q^{\frac{n}{2}}(q^{-n}-1),\quad
  b_n(\bm{\lambda})=q^{-\frac{n+1}{2}}.
\end{align}

\paragraph{annihilation/creation operators and commutation relations}
\begin{align}
  \alpha_{\pm}(\mathcal{H})&=(q^{\mp 1}-1)(\mathcal{H}+1),\qquad
  q^{\mathcal{N}}=(\mathcal{H}+1)^{-1},\\
  a^{(\pm)}&=\frac{\pm 1}{q^{-1}-q}\Bigl([\mathcal{H},\eta]_{q^{\pm 1}}
  +(1-q^{\pm 1})\bigl(\eta-\tfrac12(b_1+q^{-1}b_3)\n
  &\qquad\qquad\qquad\qquad\qquad\qquad\ \qquad
  +\tfrac12(1+q^{-1})b_3(\mathcal{H}+1)^{-1}\bigr)\Bigr)
  (\mathcal{H}+1)^{-1},\\
  b^{\text{rec}}_{n+1}-b^{\text{rec}}_{n}
  &=-\tfrac14(q^{-4}-1)qb_3^2q^{4n}+\tfrac14(q^{-3}-1)b_3(b_3+qb_1)q^{3n}\n
  &\quad
  -\tfrac14(q^{-2}-1)(b_1b_3+qb_2)q^{2n}+\tfrac14(q^{-1}-1)(b_2+q)q^n,\\
  [\mathcal{H},a^{(\pm)}]&=(q^{\mp 1}-1)a^{(\pm)}(\mathcal{H}+1),
  \label{KS3.3[H,apm]}\\
  [a^{(-)},a^{(+)}]
  &=-\tfrac14(q^{-4}-1)qb_3^2(\mathcal{H}+1)^{-4}
  +\tfrac14(q^{-3}-1)b_3(b_3+qb_1)(\mathcal{H}+1)^{-3}\n
  &\quad
  -\tfrac14(q^{-2}-1)(b_1b_3+qb_2)(\mathcal{H}+1)^{-2}
  +\tfrac14(q^{-1}-1)(b_2+q)(\mathcal{H}+1)^{-1}.
\end{align}
The r.h.s. of the above commutation relation is a quartic polynomial
in $q^{\mathcal{N}}$.
In terms of a deformed commutator we have:
\begin{equation}
  \mathcal{H}a^{(\pm)}-q^{\mp 1}a^{(\pm)}\mathcal{H}
  =(q^{\mp 1}-1)a^{(\pm)},\quad \text{namely}, \quad
  [\mathcal{H},a^{(\pm)}]_{q^{\mp 1}}=(q^{\mp 1}-1)a^{(\pm)}.
  \label{KS3.3qc}
\end{equation}
The following relation
\begin{align}
  b^{\text{rec}}_{n+1}-q^4b^{\text{rec}}_{n}
  &=-\tfrac14(1-q)b_3(b_3+qb_1)q^{3n}+\tfrac14(1-q^2)(b_1b_3+qb_2)q^{2n}\n
  &\quad
  -\tfrac14(1-q^3)(b_2+q)q^{n}+\tfrac14(1-q^4)
\end{align}
means that $[a^{(-)},a^{(+)}]_{q^4}$ is a cubic polynomial in
$q^{\mathcal N}$.

\paragraph{coherent state}
\begin{equation}
  \psi(\alpha,x\,;\bm{\lambda})=\phi_0(x\,;\bm{\lambda})
  \sum_{n=0}^{\infty}\frac{2^n\alpha^n}
  {(q\,;q)_n\prod_{1\leq j<k\leq 3}(a_ja_k\,;q)_n}\,
  P_n(\eta(x)\,;\bm{\lambda}).
  \label{52coh}
\end{equation}

\paragraph{orthogonality}
\begin{gather}
  \int_0^{\pi}\phi_0(x\,;\bm{\lambda})^2
  P_n(\eta\,;\bm{\lambda})P_m(\eta\,;\bm{\lambda})dx
  =2\pi\,
  \frac{1}{(q^{n+1}\,;q)_{\infty}
  \prod_{1\leq j<k\leq 3}(a_ja_kq^n\,;q)_{\infty}}
  \,\delta_{nm},\\
  \frac{1}{h_0(\bm{\lambda})}=\frac{1}{2\pi}
  (q\,;q)_{\infty}\!\!\prod_{1\leq j<k\leq 3}(a_ja_k\,;q)_{\infty}\,,\quad
  \frac{h_0(\bm{\lambda})}{h_n(\bm{\lambda})}=
  \frac{1}{(q\,;q)_n\prod_{1\leq j<k\leq 3}(a_ja_k\,;q)_n}\,.
\end{gather}

\subsection{Al-Salam-Chihara [KS3.8]}
\label{[KS3.8]}

This is a further restriction of the continuous dual $q$-Hahn
polynomial \S\ref{[KS3.3]} with $a_3=0$.
The dynamical symmetry algebra is further simplified and
$[a^{(-)},a^{(+)}]$ is a quadratic polynomial in $q^{\mathcal{N}}$.
The coherent state gives an explicit generating function with
symmetry $a_1\leftrightarrow a_2$ \eqref{alsalamcohe}.

\paragraph{parameters and potential function}
\begin{gather}
  q^{\bm{\lambda}}\eqdef(a_1,a_2),\ \bm{\delta}=(\tfrac12,\tfrac12),
  \ q;\quad \{a_1^*,a_2^*\}=\{a_1,a_2\} \ (\text{as a set}),\quad
  |a_i|<1,\quad i=1,2;\\[4pt]
  V(x\,;\bm{\lambda})\eqdef\frac{(1-a_1z)(1-a_2z)}
  {(1-z^2)(1-qz^2)},\quad z=e^{ix}.
\end{gather}

\paragraph{shape invariance and closure relation}
\begin{align}
  \mathcal{E}_n(\bm{\lambda})&=q^{-n}-1,\\
  R_1(y)&=(q^{-\frac12}-q^{\frac12})^2y',\quad y'\eqdef y+1,\quad
  R_0(y)=(q^{-\frac12}-q^{\frac12})^2y^{\prime\,2},\\
  R_{-1}(y)&=-\tfrac12(q^{-\frac12}-q^{\frac12})^2
  (a_1+a_2)y'.
\end{align}

\paragraph{eigenfunctions}
\begin{align}
  \phi_0(x\,;\bm{\lambda})&\eqdef
  \biggl|\frac{(e^{2ix}\,;q)_{\infty}}
  {\prod_{j=1}^2(a_je^{ix}\,;q)_{\infty}}\biggr|,
  \label{ASCphi0}\\[4pt]
  P_n(\eta\,;\bm{\lambda})&=Q_n(\cos x\,;a_1,a_2|q)
  \eqdef
  a_1^{-n}(a_1a_2\,;q)_n\,
  {}_3\phi_2\Bigl(\genfrac{}{}{0pt}{}{q^{-n},\,
  a_1e^{ix},\,a_1e^{-ix}}{a_1a_2,\,0}\Bigm|q\,;q\Bigr),
  \label{defASC}
\end{align}
which are symmetric under the permutations of $(a_1,a_2)$.
\begin{align}
  c_n&=2^n,\quad
  a_n^{\text{rec}}=\tfrac12(a_1+a_2)q^n,\quad
  b_n^{\text{rec}}=\tfrac14(1-q^n)(1-a_1a_2q^{n-1}),\\
  f_n(\bm{\lambda})&=q^{\frac{n}{2}}(q^{-n}-1),\quad
  b_n(\bm{\lambda})=q^{-\frac{n+1}{2}}.
\end{align}

\paragraph{annihilation/creation operators and commutation relations}
\begin{align}
  \alpha_{\pm}(\mathcal{H})&=(q^{\mp 1}-1)(\mathcal{H}+1),\qquad
  q^{\mathcal{N}}=(\mathcal{H}+1)^{-1},\\
  a^{(\pm)}&=\frac{\pm 1}{q^{-1}-q}\Bigl([\mathcal{H},\eta]_{q^{\pm 1}}
  +(1-q^{\pm 1})\bigl(\eta-\tfrac12(a_1+a_2)\bigr)\Bigr)
  (\mathcal{H}+1)^{-1},\\
  b^{\text{rec}}_{n+1}-b^{\text{rec}}_{n}
  &=\tfrac14(q^{-1}-1)\bigl(-(1+q)a_1a_2q^{2n}+(a_1a_2+q)q^n\bigr),\\
  [\mathcal{H},a^{(\pm)}]&=(q^{\mp 1}-1)a^{(\pm)}(\mathcal{H}+1),
  \label{KS3.8[H,apm]}\\
  [a^{(-)},a^{(+)}]
  &=\tfrac14(q^{-1}-1)\bigl(-(1+q)a_1a_2(\mathcal{H}+1)^{-2}
  +(a_1a_2+q)(\mathcal{H}+1)^{-1}\bigr).
\end{align}
The r.h.s. is a quadratic polynomial in $q^{\mathcal{N}}$.
The deformed commutators are:
\begin{equation}
  \mathcal{H}a^{(\pm)}-q^{\mp 1}a^{(\pm)}\mathcal{H}
  =(q^{\mp 1}-1)a^{(\pm)},\quad
  \text{namely},\quad
  [\mathcal{H},a^{(\pm)}]_{q^{\mp 1}}=(q^{\mp 1}-1)a^{(\pm)}.
  \label{KS3.8qc}
\end{equation}
Other interesting quantities are:
\begin{align}
  b^{\text{rec}}_{n+1}-qb^{\text{rec}}_{n}
  &=\tfrac14(1-q)(1-a_1a_2q^{2n}),
  \label{KS3.8bb1}\\
  b^{\text{rec}}_{n+1}-q^2b^{\text{rec}}_{n}
  &=\tfrac14(1-q)\bigl(1+q-(a_1a_2+q)q^n\bigr).
  \label{KS3.8bb2}
\end{align}
These mean that $[a^{(-)},a^{(+)}]_{q}$ and
$[a^{(-)},a^{(+)}]_{q^2}$ take simple forms and, in particular,
the latter is linear in $q^{\mathcal{N}}$. As we will see in another
example, the continuous $q$-Laguerre \S\ref{[KS3.19]},
  these are special to the
restricted Askey-Wilson polynomials with  a quadratic
polynomial $(1-a_1z)(1-a_2z)$ in the numerator of the potential
function $V(x)$, see
\eqref{KS3.19bb1}--\eqref{KS3.19bb2}.

\paragraph{coherent state}
\begin{align}
  \psi(\alpha,x\,;\bm{\lambda})&=\phi_0(x\,;\bm{\lambda})
  \sum_{n=0}^{\infty}\frac{2^n\alpha^n}{(q,a_1a_2\,;q)_n}\,
  P_n(\eta(x)\,;\bm{\lambda})\n
  &=\phi_0(x\,;\bm{\lambda})\,\frac{1}{(2\alpha e^{ix}\,;q)_{\infty}}\,
  {}_2\phi_1\Bigl(\genfrac{}{}{0pt}{}{a_1e^{ix},\,a_2e^{ix}}{a_1a_2}
  \Bigm|q\,;2\alpha e^{-ix}\Bigr),
  \label{alsalamcohe}
\end{align}
which is obviously symmetric in $a_1\leftrightarrow a_2$ and listed
in \cite{koeswart}.

\paragraph{orthogonality}
\begin{gather}
  \int_0^{\pi}\phi_0(x\,;\bm{\lambda})^2
  P_n(\eta\,;\bm{\lambda})P_m(\eta\,;\bm{\lambda})dx
  =2\pi\,
  \frac{1}{(q^{n+1},a_1a_2q^n\,;q)_{\infty}}
  \,\delta_{nm},\\
  \frac{1}{h_0(\bm{\lambda})}=\frac{1}{2\pi}
  (q,a_1a_2\,;q)_{\infty}\,,\quad
  \frac{h_0(\bm{\lambda})}{h_n(\bm{\lambda})}=
  \frac{1}{(q\,;q)_n(a_1a_2\,;q)_n}\,.
\end{gather}

\subsection{continuous big $q$-Hermite [KS3.18]}
\label{[KS3.18]}

This is a further restriction of the Al-Salam-Chihara polynomial
\S\ref{[KS3.8]} with $a_2=0$.
The continuous big $q$-Hermite gives another simple
realisation of the $q$-oscillator algebra \eqref{bigqosci}.

\paragraph{parameters and potential function}
\begin{gather}
  q^{\bm{\lambda}}\eqdef a,\quad \bm{\delta}=\tfrac12,\quad q;
  \quad -1<a<1;\\
  V(x\,;\bm{\lambda})\eqdef\frac{1-az}{(1-z^2)(1-qz^2)},\quad z=e^{ix}.
\end{gather}

\paragraph{shape invariance and closure relation}
\begin{align}
  \mathcal{E}_n(\bm{\lambda})&=q^{-n}-1,\\
  R_1(y)&=(q^{-\frac12}-q^{\frac12})^2y',\quad y'\eqdef y+1,\quad
  R_0(y)=(q^{-\frac12}-q^{\frac12})^2y^{\prime\,2},\\
  R_{-1}(y)&=-\tfrac12(q^{-\frac12}-q^{\frac12})^2 ay'.
\end{align}

\paragraph{eigenfunctions}
\begin{align}
  \phi_0(x\,;\bm{\lambda})&\eqdef
  \biggl|\frac{(e^{2ix}\,;q)_{\infty}}
  {(ae^{ix}\,;q)_{\infty}}\biggr|,
  \label{bqHphi0}\\
  P_n(\eta\,;\bm{\lambda})&= H_n(\cos x\,;a|q)
  \eqdef a^{-n}\,
  {}_3\phi_2\Bigl(\genfrac{}{}{0pt}{}{q^{-n},\,
  ae^{ix},\,ae^{-ix}}{0,\,0}\Bigm|q\,;q\Bigr),
  \label{defcbqH}\\
  c_n&=2^n,\quad
  a_n^{\text{rec}}=\tfrac12aq^n,\quad
  b_n^{\text{rec}}=\tfrac14(1-q^n),\\
  f_n(\bm{\lambda})&=q^{\frac{n}{2}}(q^{-n}-1),\quad
  b_n(\bm{\lambda})=q^{-\frac{n+1}{2}}.
\end{align}

\paragraph{annihilation/creation operators and commutation relations}
\begin{align}
  \alpha_{\pm}(\mathcal{H})&=(q^{\mp 1}-1)(\mathcal{H}+1),
  \qquad q^{\mathcal{N}}=(\mathcal{H}+1)^{-1},\\
  a^{(\pm)}&=\frac{\pm 1}{q^{-1}-q}\Bigl([\mathcal{H},\eta]_{q^{\pm 1}}
  +(1-q^{\pm 1})(\eta-\tfrac12a)\Bigr)
  (\mathcal{H}+1)^{-1},\\
  b^{\text{rec}}_{n+1}-b^{\text{rec}}_{n}&=\tfrac14(1-q)q^n,\\
  [\mathcal{H},a^{(\pm)}]&=(q^{\mp 1}-1)a^{(\pm)}(\mathcal{H}+1),
  \label{KS3.18[H,apm]}\\
  [a^{(-)},a^{(+)}]
  &=\tfrac14(1-q)(\mathcal{H}+1)^{-1}.
\end{align}
The deformed commutator makes \eqref{KS3.18[H,apm]} simpler
\begin{equation}
  \mathcal{H}a^{(\pm)}-q^{\mp 1}a^{(\pm)}\mathcal{H}
  =(q^{\mp 1}-1)a^{(\pm)},\quad
  \text{namely},\quad
  [\mathcal{H},a^{(\pm)}]_{q^{\mp 1}}=(q^{\mp 1}-1)a^{(\pm)}.
  \label{KS3.18qc}
\end{equation}
The relation
\begin{equation}
  b^{\text{rec}}_{n+1}-qb^{\text{rec}}_{n}=\tfrac14(1-q),
\end{equation}
implies another realisation of the $q$-oscillator
\begin{equation}
  a^{(-)}a^{(+)}-qa^{(+)}a^{(-)}
  =\tfrac14(1-q),\quad
  \text{namely},\quad
  [a^{(-)},a^{(+)}]_q=\tfrac14(1-q).
  \label{bigqosci}
\end{equation}

\paragraph{coherent state}
\eqref{psi} reads with the help of [KS(3.18.13)]
\begin{equation}
  \psi(\alpha,x\,;\bm{\lambda})=\phi_0(x\,;\bm{\lambda})
  \sum_{n=0}^{\infty}\frac{2^n\alpha^n}{(q\,;q)_n}\,
  P_n(\eta(x)\,;\bm{\lambda})
  =\phi_0(x\,;\bm{\lambda})\,\frac{(2\alpha a\,;q)_{\infty}}
  {(2\alpha e^{ix},2\alpha e^{-ix}\,;q)_{\infty}}.
  \label{54coh}
\end{equation}

\paragraph{orthogonality}
\begin{gather}
  \int_0^{\pi}\phi_0(x\,;\bm{\lambda})^2
  P_n(\eta\,;\bm{\lambda})P_m(\eta\,;\bm{\lambda})dx
  =2\pi\,
  \frac{1}{(q^{n+1}\,;q)_{\infty}}
  \,\delta_{nm},\\
  \frac{1}{h_0(\bm{\lambda})}=\frac{1}{2\pi}
  (q\,;q)_{\infty}\,,\quad
  \frac{h_0(\bm{\lambda})}{h_n(\bm{\lambda})}=
  \frac{1}{(q\,;q)_n}\,.
\end{gather}

\subsection{continuous $q$-Hermite [KS3.26]}
\label{[KS3.26]}

The continuous $q$-Hermite polynomial has been discussed in some
detail in \cite{os11} as the simplest dynamical system realising
the $q$-oscillator algebra in two different ways \eqref{qosci1}
and \eqref{qosci2}.
Here we recapitulate some formulas to make this paper complete.
Like the Hermite polynomial,  the continuous $q$-Hermite has no parameter
other than $q$.

\paragraph{parameters and potential function}
\begin{equation}
  V(x\,;\bm{\lambda})
  \eqdef\frac{1}{(1-z^2)(1-qz^2)},\quad z=e^{ix}.
\end{equation}

\paragraph{shape invariance and closure relation}
\begin{align}
  \mathcal{E}_n(\bm{\lambda})&=q^{-n}-1,\\
  R_1(y)&=(q^{-\frac12}-q^{\frac12})^2y',\quad y'\eqdef y+1,\quad
  R_0(y)=(q^{-\frac12}-q^{\frac12})^2y^{\prime\,2},\quad
  R_{-1}(y)=0.
\end{align}

\paragraph{eigenfunctions}
\begin{align}
  \phi_0(x\,;\bm{\lambda})&\eqdef\bigl|(e^{2ix}\,;q)_{\infty}\bigr|,
  \label{qHphi0}\\
  P_n(\eta\,;\bm{\lambda})&= H_n(\cos x|q)\eqdef
  e^{inx}\,
  {}_2\phi_0\Bigl(\genfrac{}{}{0pt}{}{q^{-n},\,0}{-}
  \Bigm|q\,;q^ne^{-2ix}\Bigr),
  \label{defcqHe}\\
  c_n&=2^n,\quad
  a_n^{\text{rec}}=0,\quad
  b_n^{\text{rec}}=\tfrac14(1-q^n),\\
  f_n(\bm{\lambda})&=q^{\frac{n}{2}}(q^{-n}-1),\quad
  b_n(\bm{\lambda})=q^{-\frac{n+1}{2}}.
\end{align}

\paragraph{annihilation/creation operators and commutation relations}
\begin{align}
  \alpha_{\pm}(\mathcal{H})&=(q^{\mp 1}-1)(\mathcal{H}+1),\quad
  q^{\mathcal{N}}=(\mathcal{H}+1)^{-1},\\
  a^{(\pm)}&=\frac{\pm 1}{q^{-1}-q}\Bigl([\mathcal{H},\eta]_{q^{\pm 1}}
  +(1-q^{\pm 1})\eta\Bigr)
  (\mathcal{H}+1)^{-1},\\
  b^{\text{rec}}_{n+1}-b^{\text{rec}}_{n}&=\tfrac14(1-q)q^n,\\
  [\mathcal{H},a^{(\pm)}]&=(q^{\mp 1}-1)a^{(\pm)}(\mathcal{H}+1),
  \label{KS3.26[H,apm]}\\
  [a^{(-)},a^{(+)}]
  &=\tfrac14(1-q)(\mathcal{H}+1)^{-1}.
\end{align}
The formula \eqref{KS3.26[H,apm]} can be simplified as a deformed
commutator:
\begin{equation}
  \mathcal{H}a^{(\pm)}-q^{\mp 1}a^{(\pm)}\mathcal{H}
  =(q^{\mp 1}-1)a^{(\pm)},\quad
  \text{namely},\quad
  [\mathcal{H},a^{(\pm)}]_{q^{\mp 1}}=(q^{\mp 1}-1)a^{(\pm)}.
  \label{KS3.26qc}
\end{equation}
The following relation
\begin{equation}
  b^{\text{rec}}_{n+1}-qb^{\text{rec}}_{n}=\tfrac14(1-q),
\end{equation}
means a $q$-oscillator algebra:
\begin{equation}
  a^{(-)}a^{(+)}-qa^{(+)}a^{(-)}
  =\tfrac14(1-q),\quad
  \text{namely},\quad
  [a^{(-)},a^{(+)}]_q=\tfrac14(1-q).
  \label{qosci1}
\end{equation}

\paragraph{$\bm{\lambda}$-shift operators}
Since the theory has no parameter $\bm{\lambda}$,
$\mathcal{A}^{\dagger}$ and $\mathcal{A}$ work as the creation
and annihilation operators. Thus $a^{(+)}$ and $\mathcal{A}^{\dagger}$
($a^{(-)}$ and $\mathcal{A}$) are closely related:
\begin{align}
  a^{(+)}&=\mathcal{A}^{\dagger}X,\\
  \mathcal{A}^{\dagger}&=-i\bigl(\sqrt{V(x)}\,e^{\gamma p/2}
  -\sqrt{V(x)^*}\,e^{-\gamma p/2}\bigr),\\
  X&\eqdef -\frac{i}{2}q\bigl(z\sqrt{V(x)}\,e^{\gamma p/2}
  -z^{-1}\sqrt{V(x)^*}\,e^{-\gamma p/2}\bigr)(\mathcal{H}+1)^{-1}.
\end{align}
The similarity transformed quantities are:
\begin{align}
  \widetilde{a}^{(+)}&=\widetilde{\mathcal{A}}^{\dagger}\widetilde{X},\\
  \widetilde{\mathcal{A}}^{\dagger}&\eqdef
  \phi_0(x)^{-1}\circ\mathcal{A}^{\dagger}\circ\phi_0(x)
  =q^{-\frac12}\Bigl(
  \frac{z^{-1}}{1-z^2}\,e^{\gamma p/2}
  +\frac{z}{1-z^{-2}}\,e^{-\gamma p/2}\Bigr),\\
  \widetilde{X}&\eqdef\phi_0(x)^{-1}\circ X\circ\phi_0(x)
  =\frac12q^{\frac12}\Bigl(\frac{1}{1-z^2}\,e^{\gamma p/2}
  +\frac{1}{1-z^{-2}}\,e^{-\gamma p/2}\Bigr)(\widetilde{\mathcal{H}}+1)^{-1}.
\end{align}
As there is no $\bm{\lambda}$ to be shifted, we have
$\widetilde{\mathcal{A}}^{\dagger}=\mathcal{B}(\bm{\lambda})$
\eqref{Bshift} and $\widetilde{\mathcal{A}}=\mathcal{F}(\bm{\lambda})$
\eqref{Fshift}.
The $\widetilde{X}$ and $\widetilde{A}^{\dagger}$ operators work as
\begin{align}
  \widetilde{X}P_n(x)=\frac12q^{\frac{n+1}{2}}P_n(x),\qquad
  \widetilde{\mathcal{A}}^{\dagger}P_n(x)=q^{-\frac{n+1}{2}}P_{n+1}(x)
\end{align}
and $\widetilde{X}$ satisfies the relation
\begin{align}
  \Bigl(2q^{-\frac12}\widetilde{X}(\widetilde{\mathcal{H}}+1)\Bigr)^2
  &=\Bigl(\frac{1}{1-z^2}\,e^{\gamma p/2}
  +\frac{1}{1-z^{-2}}\,e^{-\gamma p/2}\Bigr)^2\n
  &=V(x)e^{\gamma p}+V(x)^*e^{-\gamma p}-V(x)-V(x)^*+1\n
  &=\widetilde{\mathcal{H}}+1.
\end{align}
It is easy to verify that the shape invariance relation \eqref{shapeinv}
itself implies a realisation of the $q$-oscillator algebra with
$\mathcal{A}$ and $\mathcal{A}^{\dagger}$ \cite{os11}:
\begin{equation}
  \mathcal{A}\mathcal{A}^{\dagger}-q^{-1}\mathcal{A}^{\dagger}\mathcal{A}
  =q^{-1}-1,\quad
  \text{namely},\quad
  [\mathcal{A},\mathcal{A}^{\dagger}]_{q^{-1}}=q^{-1}-1.
  \label{qosci2}
\end{equation}

\paragraph{coherent state}
\eqref{psi} reads with the help of [KS(3.26.11)]
\begin{equation}
  \psi(\alpha,x\,;\bm{\lambda})=\phi_0(x\,;\bm{\lambda})
  \sum_{n=0}^{\infty}\frac{2^n\alpha^n}{(q\,;q)_n}\,
  P_n(\eta(x)\,;\bm{\lambda})
  =\phi_0(x\,;\bm{\lambda})\,
  \frac{1}{(2\alpha e^{ix},2\alpha e^{-ix}\,;q)_{\infty}}.
  \label{55coh}
\end{equation}

\paragraph{orthogonality}
\begin{gather}
  \int_0^{\pi}\phi_0(x\,;\bm{\lambda})^2
  P_n(\eta\,;\bm{\lambda})P_m(\eta\,;\bm{\lambda})dx
  =2\pi\,
  \frac{1}{(q^{n+1}\,;q)_{\infty}}
  \,\delta_{nm},\\
  \frac{1}{h_0(\bm{\lambda})}=\frac{1}{2\pi}
  (q\,;q)_{\infty}\,,\quad
  \frac{h_0(\bm{\lambda})}{h_n(\bm{\lambda})}=
  \frac{1}{(q\,;q)_n}\,.
\end{gather}

\subsection{continuous $q$-Jacobi [KS3.10]}
\label{[KS3.10]}

\paragraph{parameters and potential function}
\begin{gather}
  \bm{\lambda}\eqdef(\alpha,\beta),\quad \bm{\delta}=(1,1),\quad q;\quad
  \alpha,\beta\ge -\frac{1}{2};\\
  V(x\,;\bm{\lambda})\eqdef
  \frac{(1-q^{\frac12(\alpha+\frac12)}z)(1-q^{\frac12(\alpha+\frac32)}z)
  (1+q^{\frac12(\beta+\frac12)}z)(1+q^{\frac12(\beta+\frac32)}z)}
  {(1-z^2)(1-qz^2)},\quad z=e^{ix}.
\end{gather}

\paragraph{shape invariance and closure relation}
\begin{align}
  \mathcal{E}_n(\bm{\lambda})&=(q^{-n}-1)(1-q^{n+\alpha+\beta+1}),\\
  R_1(y)&=(q^{-\frac12}-q^{\frac12})^2y',\quad
  y'\eqdef y+1+q^{\alpha+\beta+1},\\
  R_0(y)&=(q^{-\frac12}-q^{\frac12})^2\bigl(
  y^{\prime\,2}-(1+q)^2q^{\alpha+\beta}\bigr),\\
  R_{-1}(y)&=-\tfrac12(q^{-\frac12}-q^{\frac12})^2
  q^{\frac14}(1+q^{\frac12})(q^{\frac12\alpha}-q^{\frac12\beta})
  (1-q^{\frac12(\alpha+\beta)})
  \bigl(y'+(1+q)q^{\frac12(\alpha+\beta)}\bigr).
\end{align}

\paragraph{eigenfunctions}
\begin{align}
  \phi_0(x\,;\bm{\lambda})&\eqdef
  \biggl|\frac{(e^{2ix}\,;q)_{\infty}}
  {(q^{\frac12(\alpha+\frac12)}e^{ix},
   -q^{\frac12(\beta+\frac12)}e^{ix}\,;q^{\frac12})_{\infty}}\biggr|,
  \label{qJphi0}\\
  P_n(\eta\,;\bm{\lambda})&= P_n^{(\alpha,\beta)}(\cos x|q)\n
  &\eqdef
  \frac{(q^{\alpha+1}\,;q)_n}{(q\,;q)_n}\,
  {}_4\phi_3\Bigl(\genfrac{}{}{0pt}{}{q^{-n},\,q^{n+\alpha+\beta+1},\,
  q^{\frac12(\alpha+\frac12)}e^{ix},\,q^{\frac12(\alpha+\frac12)}e^{-ix}}
  {q^{\alpha+1},\,-q^{\frac12(\alpha+\beta+1)},\,
  -q^{\frac12(\alpha+\beta+2)}}\Bigm|q\,;q\Bigr),
  \label{defcqJ}
\end{align}
\begin{align}
  c_n&=\frac{2^nq^{\frac12(\alpha+\frac12)n}(q^{n+\alpha+\beta+1}\,;q)_n}
  {(q,-q^{\frac12(\alpha+\beta+1)},-q^{\frac12(\alpha+\beta+2)}\,;q)_n},\\
  a_n^{\text{rec}}&=\frac12\biggl(q^{\frac12(\alpha+\frac12)}
  +q^{-\frac12(\alpha+\frac12)}\n
  &\qquad
  -\frac{(1-q^{n+\alpha+1})(1-q^{n+\alpha+\beta+1})
  (1+q^{n+\frac12(\alpha+\beta+1)})(1+q^{n+\frac12(\alpha+\beta+2)})}
  {q^{\frac12(\alpha+\frac12)}(1-q^{2n+\alpha+\beta+1})
  (1-q^{2n+\alpha+\beta+2})}\n
  &\qquad
  -\frac{q^{\frac12(\alpha+\frac12)}(1-q^n)(1-q^{n+\beta})
  (1+q^{n+\frac12(\alpha+\beta)})(1+q^{n+\frac12(\alpha+\beta+1)})}
  {(1-q^{2n+\alpha+\beta})(1-q^{2n+\alpha+\beta+1})}\biggr),\\
  b_n^{\text{rec}}&=
  (1-q^n)(1-q^{n+\alpha})(1-q^{n+\beta})(1-q^{n+\alpha+\beta})\n
  &\quad\times
  \frac{(1+q^{n+\frac12(\alpha+\beta-1)})
  (1+q^{n+\frac12(\alpha+\beta)})^2
  (1+q^{n+\frac12(\alpha+\beta+1)})}
  {4(1-q^{2n+\alpha+\beta-1})(1-q^{2n+\alpha+\beta})^2
  (1-q^{2n+\alpha+\beta+1})},\\
  f_n(\bm{\lambda})&=\frac{q^{\frac12(\alpha+\frac32)}q^{-n}
  (1-q^{n+\alpha+\beta+1})}
  {(1+q^{\frac12(\alpha+\beta+1)})(1+q^{\frac12(\alpha+\beta+2)})},\\
  b_n(\bm{\lambda})&=q^{-\frac12(\alpha+\frac32)}q^{n+1}(q^{-(n+1)}-1)
  (1+q^{\frac12(\alpha+\beta+1)})(1+q^{\frac12(\alpha+\beta+2)}).
\end{align}

\paragraph{annihilation/creation operators and commutation relations}
\begin{align}
  \alpha_{\pm}(\mathcal{H})
  &=\tfrac12(q^{-\frac12}-q^{\frac12})^2\mathcal{H}'
  \pm\tfrac12(q^{-1}-q)\sqrt{\mathcal{H}^{\prime\,2}-4q^{\alpha+\beta+1}},
  \quad\,
  \mathcal{H}'\eqdef\mathcal{H}+1+q^{\alpha+\beta+1},\\
  q^{\mathcal{N}}&=\tfrac12q^{-\alpha-\beta-1}\bigl(\mathcal{H}'
  -\sqrt{\mathcal{H}^{\prime\,2}-4q^{\alpha+\beta+1}}\,\bigr)\quad
  (\text{for }0<q^{\alpha+\beta+1}<1),\\
  [\mathcal{H},a^{(\pm)}]&=\tfrac12a^{(\pm)}\Bigl(
  (q^{-\frac12}-q^{\frac12})^2\mathcal{H}'
  \pm(q^{-1}-q)\sqrt{\mathcal{H}^{\prime\,2}-4q^{\alpha+\beta+1}}\Bigr).
\end{align}
The annihilation/creation operators \eqref{a^{(pm)}} and their
commutation relation \eqref{[a-,a+]} are not so simplified because
$b^{\text{rec}}_{n+1}-b^{\text{rec}}_{n}
=q^n\times(\text{9-th degree polynomial in $q^n$})/
(\text{11-th degree polynomial}$ $\text{in $q^n$})$
has a lengthy expression.

\paragraph{coherent state}
We are not aware if a simple summation formula exists for the coherent
state:
\begin{equation}
  \psi(\alpha',x\,;\bm{\lambda})=\phi_0(x\,;\bm{\lambda})
  \sum_{n=0}^{\infty}\frac{(2q^{-\frac12(\alpha+\frac12)})^n
  (q^{\frac12(\alpha+\beta)+1}\,;q^{\frac12})_{2n}\alpha^{\prime\,n}}
  {(q^{\alpha+1},q^{\beta+1}\,;q)_n}\,
  P_n(\eta(x)\,;\bm{\lambda}).
  \label{56coh}
\end{equation}

\paragraph{orthogonality}
\begin{align}
  &\int_0^{\pi}\phi_0(x\,;\bm{\lambda})^2
  P_n(\eta\,;\bm{\lambda})P_m(\eta\,;\bm{\lambda})dx\n
  =\,&2\pi\,
  \frac{(1-q^{\alpha+\beta+1})
   (q^{\alpha+1},q^{\beta+1},-q^{\frac12(\alpha+\beta+3)}\,;q)_n}
  {(1-q^{2n+\alpha+\beta+1})
   (q,q^{\alpha+\beta+1},-q^{\frac12(\alpha+\beta+1)}\,;q)_n}\,
  q^{(\alpha+\frac12)n}\n
  &\quad\times
  \frac{(q^{\frac12(\alpha+\beta+2)},
  q^{\frac12(\alpha+\beta+3)}\,;q)_{\infty}}
  {(q,q^{\alpha+1},q^{\beta+1},-q^{\frac12(\alpha+\beta+1)},
    -q^{\frac12(\alpha+\beta+2)}\,;q)_{\infty}}
  \,\delta_{nm},\\[4pt]
  \frac{1}{h_0(\bm{\lambda})}&=
  \frac{(q,q^{\alpha+1},q^{\beta+1},-q^{\frac12(\alpha+\beta+1)},
         -q^{\frac12(\alpha+\beta+2)}\,;q)_{\infty}}
       {2\pi(q^{\frac12(\alpha+\beta+2)},
        q^{\frac12(\alpha+\beta+3)}\,;q)_{\infty}}\,,\\
  \frac{h_0(\bm{\lambda})}{h_n(\bm{\lambda})}&=
  \frac{(1-q^{2n+\alpha+\beta+1})
   (q,q^{\alpha+\beta+1},-q^{\frac12(\alpha+\beta+1)}\,;q)_n}
  {(1-q^{\alpha+\beta+1})
   (q^{\alpha+1},q^{\beta+1},-q^{\frac12(\alpha+\beta+3)}\,;q)_n}\,
  q^{-(\alpha+\frac12)n}.
\end{align}

\subsection{continuous $q$-Laguerre [KS3.19]}
\label{[KS3.19]}

This is a further restriction ($\beta\to\infty$ or $q^\beta\to0$) of 
the continuous $q$-Jacobi polynomial \S\ref{[KS3.10]}.
Many formulas are drastically simplified.

\paragraph{parameters and potential function}
\begin{gather}
  \bm{\lambda}\eqdef\alpha,\quad \bm{\delta}=1,\quad q;
  \quad \alpha\geq-\frac12;\\
  V(x\,;\bm{\lambda})\eqdef
  \frac{(1-q^{\frac12(\alpha+\frac12)}z)(1-q^{\frac12(\alpha+\frac32)}z)}
  {(1-z^2)(1-qz^2)},\quad z=e^{ix}.
\end{gather}

\paragraph{shape invariance and closure relation}
\begin{align}
  \mathcal{E}_n(\bm{\lambda})&=q^{-n}-1,\\
  R_1(y)&=(q^{-\frac12}-q^{\frac12})^2y',\quad y'\eqdef y+1,\quad
  R_0(y)=(q^{-\frac12}-q^{\frac12})^2y^{\prime\,2},\\
  R_{-1}(y)&=-\tfrac12(q^{-\frac12}-q^{\frac12})^2
  q^{\frac12(\alpha+\frac12)}(1+q^{\frac12})y'.
\end{align}

\paragraph{eigenfunction}
\begin{align}
  \phi_0(x\,;\bm{\lambda})&\eqdef
  \biggl|\frac{(e^{2ix}\,;q)_{\infty}}
  {(q^{\frac12(\alpha+\frac12)}e^{ix}\,;q^{\frac12})_{\infty}}\biggr|,
  \label{qLphi0}\\
  P_n(\eta\,;\bm{\lambda})&=P_n^{(\alpha)}(\cos x|q)\eqdef
  \frac{(q^{\alpha+1}\,;q)_n}{(q\,;q)_n}\,
  {}_3\phi_2\Bigl(\genfrac{}{}{0pt}{}{q^{-n},\,
  q^{\frac12(\alpha+\frac12)}e^{ix},\,q^{\frac12(\alpha+\frac12)}e^{-ix}}
  {q^{\alpha+1},\,0}\Bigm|q\,;q\Bigr),
  \label{defcqL}\\
  c_n&=\frac{2^nq^{\frac12(\alpha+\frac12)n}}{(q\,;q)_n},\quad
  a_n^{\text{rec}}=\tfrac12q^{n+\frac12(\alpha+\frac12)}(1+q^{\frac12}),\quad
  b_n^{\text{rec}}=\tfrac14(1-q^n)(1-q^{n+\alpha}),\\
  f_n(\bm{\lambda})&=q^{\frac12(\alpha+\frac32)}q^{-n},\quad
  b_n(\bm{\lambda})=q^{-\frac12(\alpha+\frac32)}q^{n+1}(q^{-(n+1)}-1).
\end{align}

\paragraph{annihilation/creation operators and commutation relations}
\begin{align}
  \alpha_{\pm}(\mathcal{H})&=(q^{\mp 1}-1)(\mathcal{H}+1),\qquad
  q^{\mathcal{N}}=(\mathcal{H}+1)^{-1},\\
  a^{(\pm)}&=\frac{\pm 1}{q^{-1}-q}\Bigl([\mathcal{H},\eta]_{q^{\pm 1}}
  +(1-q^{\pm 1})\bigl(\eta-\tfrac12q^{\frac12(\alpha+\frac12)}
  (1+q^{\frac12})\bigr)\Bigr)
  (\mathcal{H}+1)^{-1},\\
  b^{\text{rec}}_{n+1}-b^{\text{rec}}_{n}
  &=\tfrac14(1-q)\bigl(-(1+q)q^{\alpha}q^{2n}+(1+q^{\alpha})q^n\bigr),\\
  [\mathcal{H},a^{(\pm)}]&=(q^{\mp 1}-1)a^{(\pm)}(\mathcal{H}+1),
  \label{KS3.19[H,apm]}\\
  [a^{(-)},a^{(+)}]
  &=\tfrac14(1-q)\bigl(-(1+q)q^{\alpha}(\mathcal{H}+1)^{-2}
  +(1+q^{\alpha})(\mathcal{H}+1)^{-1}\bigr).
\end{align}
Again \eqref{KS3.19[H,apm]} can be written as $q$-deformed commutators:
\begin{equation}
  \mathcal{H}a^{(\pm)}-q^{\mp 1}a^{(\pm)}\mathcal{H}
  =(q^{\mp 1}-1)a^{(\pm)},\quad
  \text{namely},\quad
  [\mathcal{H},a^{(\pm)}]_{q^{\mp 1}}=(q^{\mp 1}-1)a^{(\pm)}.
  \label{KS3.19qc}
\end{equation}
The following mean that $[a^{(-)},a^{(+)}]_{q}$ and
$[a^{(-)},a^{(+)}]_{q^2}$ take simple forms, see
\eqref{KS3.8bb1}--\eqref{KS3.8bb2}:
\begin{align}
  b^{\text{rec}}_{n+1}-qb^{\text{rec}}_{n}
  &=\tfrac14(1-q)(1-q^{\alpha+1+2n}),
  \label{KS3.19bb1}\\
  b^{\text{rec}}_{n+1}-q^2b^{\text{rec}}_{n}
  &=\tfrac14(1-q)\bigl(1+q-(1+q^{\alpha})q^{n+1}\bigr).
  \label{KS3.19bb2}
\end{align}

\paragraph{coherent state}
\eqref{psi} reads with the help of [KS(3.19.12)]
\begin{align}
  \psi(\alpha',x\,;\bm{\lambda})&=\phi_0(x\,;\bm{\lambda})
  \sum_{n=0}^{\infty}\frac{(2q^{-\frac12(\alpha+\frac12)})^n
  \alpha^{\prime\,n}}{(q^{\alpha+1}\,;q)_n}\,
  P_n(\eta(x)\,;\bm{\lambda})\n
  &=\phi_0(x\,;\bm{\lambda})\,\frac{1}{(2\alpha' e^{ix}\,;q)_{\infty}}\,
  {}_2\phi_1\Bigl(\genfrac{}{}{0pt}{}
  {q^{\frac12(\alpha+\frac12)}e^{ix},\,q^{\frac12(\alpha+\frac32)}e^{ix}}
  {q^{\alpha+1}}\Bigm|q\,;2\alpha' e^{-ix}\Bigr).
  \label{57coh}
\end{align}

\paragraph{orthogonality}
\begin{gather}
  \int_0^{\pi}\phi_0(x\,;\bm{\lambda})^2
  P_n(\eta\,;\bm{\lambda})P_m(\eta\,;\bm{\lambda})dx
  =2\pi\,
  \frac{(q^{\alpha+1}\,;q)_n}{(q\,;q)_n}\,q^{(\alpha+\frac12)n}
  \frac{1}{(q,q^{\alpha+1}\,;q)_{\infty}}
  \,\delta_{nm},\\
  \frac{1}{h_0(\bm{\lambda})}=\frac{1}{2\pi}
  (q,q^{\alpha+1}\,;q)_{\infty}\,,\quad
  \frac{h_0(\bm{\lambda})}{h_n(\bm{\lambda})}=
  \frac{(q\,;q)_n}{(q^{\alpha+1}\,;q)_n}\,
  q^{-(\alpha+\frac12)n}.
\end{gather}

\subsection{Comments on the two polynomials with $\eta(x)=\cos(x+\phi)$}
\label{seccosxp}

In the review of Koekoek and Swarttouw \cite{koeswart}, two polynomials,
the continuous $q$-Hahn [KS3.4] and the $q$-Meixner-Pollaczek [KS3.9]
are listed as having $\eta(x)=\cos(x+\phi)$, with non-vanishing angle
$\phi$ appearing in the definition of polynomials.
In fact, the continuous $q$-Hahn polynomial is the same as the
Askey-Wilson polynomial \S\ref{[KS3.1]} and the $q$-Meixner-Pollaczek
polynomial is proportional to the Al-Salam-Chihara polynomial \S\ref{[KS3.8]}
with  degree- or $n$- dependent coefficients.
Therefore we will not treat them as independent `discrete' quantum
mechanical systems.

\subsubsection{continuous $q$-Hahn [KS3.4]}
\label{[KS3.4]}

A simple comparison of the normalised three term recurrence relation
for the continuous $q$-Hahn polynomial (KS3.4.4) with that for
the Askey-Wilson polynomial (KS3.1.5) reveals that they are one
and the same polynomial after the identification of the parameters
(in the notation of \cite{koeswart})
\begin{equation}
  a^{\text{AW}}\to a\,e^{i\phi},\quad b^{\text{AW}}\to b\,e^{i\phi},\quad
  c^{\text{AW}}\to c\,e^{-i\phi},\quad d^{\text{AW}}\to d\,e^{-i\phi},
\end{equation}
in which the superscript {\scriptsize AW} indicates the quantity of the
Askey-Wilson polynomial.

\subsubsection{$q$-Meixner-Pollaczek [KS3.9]}
\label{[KS3.9]}

Likewise, the normalised three term recurrence relation for the
$q$-Meixner-Pollaczek polynomial $P_n^{q\text{MP}}(\eta)$ (KS3.9.4)
is the same as that for the Al-Salam-Chihara polynomial
$P_n^{\text{ASC}}(\eta)$ (KS3.8.4) after the identification
\begin{equation}
  a^{\text{ASC}}\to a\,e^{i\phi},\quad
  b^{\text{ASC}}\to a\,e^{-i\phi}; \qquad
  a^{\text{ASC}}=(b^{\text{ASC}})^*\in \mathbb{C},\quad a>0,
\end{equation}
in which the superscript {\scriptsize ASC} denotes the quantity of
the Al-Salam-Chihara polynomial. These two polynomials are different
only by a multiplicative constant:
\begin{equation}
  \frac{P_n^{\text{ASC}}(\eta)}{(q;q)_n}=P_n^{q\text{MP}}(\eta).
\end{equation}

\section{Summary and Comments}
\setcounter{equation}{0}

Known examples of exactly solvable `discrete' quantum mechanics
of one degree of freedom are discussed in detail and in full
generality.
The shape invariance property, the exact solutions in the
Schr\"odinger and Heisenberg pictures, the annihilation/creation
operators together with their symmetry algebra, the coherent state
as the eigenvector of the annihilation operator, the ground state
wavefunction giving the orthogonality measure of the eigenpolynomial
are given explicitly for each system, which is named after the
corresponding orthogonal polynomial.
The present paper supplements the earlier results
\cite{os4,os5,os6,os7,os12}. The main focus is the polynomials
obtained by restricting the Askey-Wilson polynomials. In general,
they have simple and tractable symmetry algebras, some of them are
the $q$-oscillator algebra \cite{os11}.
Another main feature is the coherent states. As many as eleven
new and exact coherent states are presented \eqref{31coh},
\eqref{32coh}, \eqref{41coh}, \eqref{42coh}, \eqref{51coh},
\eqref{52coh}, \eqref{alsalamcohe}, \eqref{54coh}, \eqref{55coh},
\eqref{56coh}, \eqref{57coh} as the eigenvectors of
the annihilation operators for the `discrete' quantum mechanical
systems. These coherent states are by construction totally symmetric
in the symmetric parameters of the Hamiltonians.
In other words, they realise the dynamically favourable generating
functions of the eigenpolynomials. Like the standard coherent state
of the harmonic oscillator, these new coherent states are expected
to find various applications in many branches of physical sciences,
in particular, quantum optics and quantum information.
It would be interesting to investigate if and to what extent these
new coherent states share the remarkable properties of the standard
coherent state of the harmonic oscillator.

One interesting future task is to solve the closure relation
\eqref{closurerel1}--\eqref{closurerel3} algebraically to determine
all the possible forms of the sinusoidal coordinate $\eta(x)$ and
the potential function $V(x)$.
For the ordinary quantum mechanics and for the orthogonal polynomials
of discrete measures, this task was done in Appendix A of \cite{os7}
and Appendix A of \cite{os12}.
The present case is more complicated than these due to the presence
of arbitrary periodic functions with period $i\gamma$.
It is interesting to see if  difference equation versions of the
soliton potential, i.e. $1/\cosh^2x$ potential in ordinary quantum
mechanics, (see, for example, \S3.1.3 of \cite{os7}) with
$\eta(x)=\sinh x$, and the Morse potential with $\eta(x)=e^{-x}$
(see \S3.1.4 of \cite{os7}) are contained as solutions or not.

\section*{Acknowledgements}

This work is supported in part by Grants-in-Aid for Scientific Research
from the Ministry of Education, Culture, Sports, Science and Technology,
No.18340061 and No.19540179.

\section*{Appendix A: Diagrammatic proof of the hermiticity
of the Hamiltonian}
\label{appendA}
\setcounter{equation}{0}
\renewcommand{\theequation}{A.\arabic{equation}}

Here we give a diagrammatic proof of the hermiticity (self-adjointness)
of the Hamiltonian \eqref{H} for the three different cases of the
sinusoidal coordinates corresponding to sections
\ref{secx}--\ref{seccosx}.
A less detailed proof of the hermiticity can be found in \cite{os10}.
The hermiticity or self-adjointness of the Hamiltonian $\mathcal{H}$
means $(g,\mathcal{H}f)=(\mathcal{H}g,f)$ for a given inner product
$(g,f)$ \eqref{inner} for arbitrary elements $f$ and $g$ of the 
appropriate Hilbert
space. It is necessary and sufficient to show that in a certain dense
subspace of the Hilbert space. The obvious choice for such a subspace
is spanned by the ground state wavefunction $\phi_0$, which is given in
each subsection \eqref{Hahnphi0}, \eqref{Meixphi0}, \eqref{Wilsonphi0},
\eqref{dualHahnphi0}, \eqref{AWilsonphi0}, \eqref{qHahnphi0},
\eqref{ASCphi0}, \eqref{bqHphi0}, \eqref{qHphi0}, \eqref{qJphi0},
\eqref{qLphi0}, times the eigenpolynomials $P_n(\eta(x))$.
The types of the polynomials are:
\begin{align}
  &(a): \text{polynomials in $\eta(x)=x$ for the Hamiltonians in
  section \ref{secx}},\n
  &\qquad (g,f)=\int_{-\infty}^{\infty}g(x)^*f(x)dx,\quad
  f(x)=\phi_0(x)P(x),\quad g(x)=\phi_0(x)Q(x),
  \label{innera1}\\
  &(b): \text{polynomials in $\eta(x)=x^2$ for the Hamiltonians in
  section \ref{secx2}},\n
  &\qquad (g,f)=\int_0^{\infty}g(x)^*f(x)dx,\quad
  f(x)=\phi_0(x)P(x^2),\quad g(x)=\phi_0(x)Q(x^2),\\
  &(c): \text{polynomials in $\eta(x)=\cos x$ for the Hamiltonians
  in section \ref{seccosx}},\n
  &\qquad (g,f)=\int_0^{\pi}g(x)^*f(x)dx,\quad
  f(x)=\phi_0(x)P(\cos x),\quad g(x)=\phi_0(x)Q(\cos x).
  \label{innera3}
\end{align}
This clearly removes the non-uniqueness of the eigenfunctions,
which was mentioned in section two.
For the Hamiltonian \eqref{factham} $\mathcal{H}=T_++T_--V(x)-V(x)^*$,
it is obvious that the function part $-V(x)-V(x)^*$ is hermitian
by itself.
When $T_+=\sqrt{V(x)}\,e^{\gamma p}\sqrt{V(x)^*}$ acts on $f$, the
argument of $f$ is shifted from $x$ to $x-i\gamma$.
With the compensating change of integration variable from $x$ to
$x+i\gamma$ one can formally show $(g,T_+f)=(T_+g,f)$ in a
straightforward way.
Similarly we have $(g,T_-f)=(T_-g,f)$ by another change of integration
variable $x$ to $x-i\gamma$.
This is the `formal hermiticity.'

\begin{figure}[t]
  \centering
  \includegraphics*[scale=1.0]{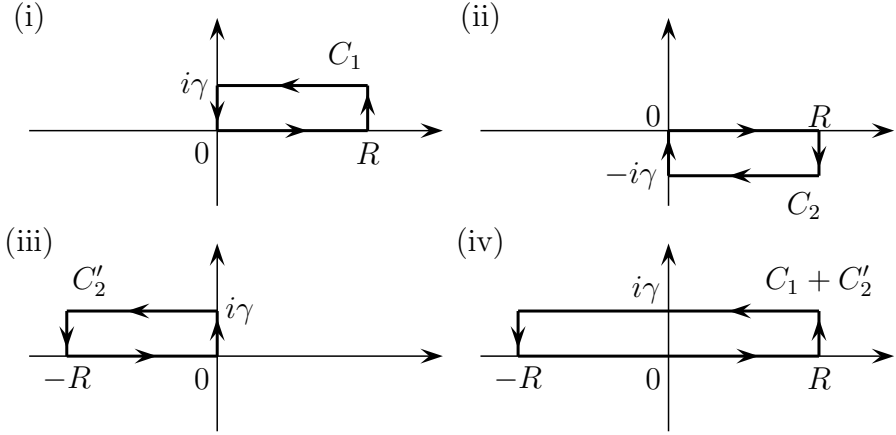}
  \caption{Integration contours in complex $x$ plane. The endpoint 
$R=\infty$ for cases ($a$)
  and ($b$), $R=\pi$ for case ($c$).
  (For case ($c$), $i\gamma$ is in the lower half plane because of
  $\gamma=\log q<0$.)}
  \label{contour}
\end{figure}

In reality, the shift of integration variable, to be realised by the
Cauchy integral, would involve additional integration contours:
\begin{alignat}{2}
  &(a): (-\infty,\pm i-\infty),\quad (+\infty,\pm i+\infty)
  &\quad&\text{for the Hamiltonians in section \ref{secx}},
  \label{1type}\\
  &(b): (0,\pm i),\quad (+\infty,\pm i+\infty)
  &\quad&\text{for the Hamiltonians in section \ref{secx2}},\\
  &(c): (0,\pm i\log q),\quad (\pi,\pi \pm i\log q)
  &\quad&\text{for the Hamiltonians in section \ref{seccosx}}.
  \label{3type}
\end{alignat}
It is easy to verify that all the singularities arising from $V$ and
$V^*$ in cases ($b$) and ($c$) are cancelled by the zeros coming from
the ground state wavefunctions $\phi_0$ and $\phi_0^*$, and the Cauchy
integration formula applies in all cases.
As can be seen from the  diagrams in Fig.\ref{contour} the contribution of the
additional contour integrals (\ref{1type})--(\ref{3type}) cancel with
each other and the shifts of integration variables are justified and
the hermiticity is established.

First, the contribution from the contours at infinity in ($a$) vanish
identically due to the strong damping by $\phi_0$ and $\phi_0^*$,
see \eqref{Hahnphi0} and \eqref{Meixphi0}. This establishes the
hermiticity in the case ($a$).
Next let us discuss the case ($b$) in detail. In this case $\gamma=1$.
The integrand of $(g,T_\pm f)$ are
\begin{align}
  g^*T_+f&=\phi_0(x)^*Q(x^2)^*\sqrt{V(x)}\sqrt{V(x+i)^*}\phi_0(x-i)P((x-i)^2)
  \eqdef F(x),\\
  g^*T_-f&=\phi_0(x)^*Q(x^2)^*\sqrt{V(x)^*}\sqrt{V(x+i)}\phi_0(x+i)P((x+i)^2)
  \eqdef G(x).
\end{align}
Due to the evenness of the eigenfunctions, $\phi_0(-x)=\phi_0(x)$,
$P((-x)^2)=P(x^2)$, $Q((-x)^2)=Q(x^2)$ and $V(x)^*=V(-x^*)$, we have
\begin{align}
  G(x)&=\phi_0(-x)^*Q((-x)^2)^*\sqrt{V(-x)}\sqrt{V(-x+i)^*}
  \phi_0(-x-i)P((-x-i)^2)\n
  &=F(-x).
  \label{FGeq}
\end{align}
On the other hand, the integrand of $(T_\pm g,f)$ are
\begin{align}
  (T_+g)^*f=\sqrt{V(x)^*}\sqrt{V(x+i)}\phi_0(x-i)^*Q((x-i)^2)^*\phi_0(x)P(x^2)
  &=F(x+i),\\
  (T_-g)^*f=\sqrt{V(x)}\sqrt{V(x+i)^*}\phi_0(x+i)^*Q((x+i)^2)^*\phi_0(x)P(x^2)
  &=G(x-i)\n
  &=F(-x+i),
\end{align}
in which \eqref{FGeq} is used for the last equality.
Since the integrands are analytic in $x$ and there is no pole within
the contours, see Fig.\ref{contour}, we have
\begin{equation}
  \oint_{C_1}F(x)dx=0,\qquad
  \oint_{C_2}G(x)dx=\oint_{C_2}F(-x)dx=\oint_{C'_2}F(x)dx=0.
\end{equation}
Combining them, we obtain
\begin{align}
  0&=\oint_{C_1}F(x)dx+\oint_{C'_2}F(x)dx=\oint_{C_1+C'_2}\!\!\!F(x)dx\n
  &=\int_{-\infty}^{\infty}F(x)dx-\int_{-\infty}^{\infty}F(x+i)dx
  +\int_{\uparrow\ \text{at}+\infty}\!\!\!\!\!F(x)dx
  +\int_{\downarrow\ \text{at}-\infty}\!\!\!\!\!F(x)dx.
  \label{x2int}
\end{align}
The contribution from the contours at infinity in the case ($b$)
vanish identically due to the strong damping by $\phi_0$ and
$\phi_0^*$, see \eqref{Wilsonphi0} and \eqref{dualHahnphi0}.
Thus \eqref{x2int} implies
$\int_{-\infty}^\infty F(x)dx=\int_{-\infty}^\infty F(x+i)dx$.
The l.h.s. is
\begin{equation}
  \int_0^{\infty}F(x)dx+\int_0^{\infty}G(x)dx=(g,T_+ f)+(g,T_- f).
\end{equation}
The r.h.s. is
\begin{equation}
  \int_0^{\infty}F(x+i)dx+\int_0^{\infty}G(x-i)dx=(T_+g,f)+(T_-g,f).
\end{equation}
Thus the hermiticity of the Hamiltonians for the case ($b$) is proved.
The hermiticity of the Hamiltonians for the case ($c$) is proved
in a similar way together with the evenness and the $2\pi$ periodicity
of the ground state wavefunction $\phi_0(x)$, the sinusoidal coordinate
$\eta(x)=\cos x$ and the potential function $V(x)$;
$\phi_0(-x)=\phi_0(x)$, $\eta(-x)=\eta(x)$, $V(x)^*=V(-x^*)$,
$\phi_0(x+2\pi)=\phi(x)$, $\eta(x+2\pi)=\eta(x)$, $V(x+2\pi)=V(x)$.

\section*{Appendix B: Some definitions related to the hypergeometric
and $q$-hypergeometric functions}
\label{appendB}
\setcounter{equation}{0}
\renewcommand{\theequation}{B.\arabic{equation}}

For self-containedness we collect several definitions related to
the ($q$-)hypergeometric functions \cite{koeswart}.

\noindent
$\circ$ Pochhammer symbol $(a)_n$ :
\begin{equation}
   (a)_n\eqdef\prod_{k=1}^n(a+k-1)=a(a+1)\cdots(a+n-1)
   =\frac{\Gamma(a+n)}{\Gamma(a)}.
   \label{defPoch}
\end{equation}
$\circ$ $q$-Pochhammer symbol $(a\,;q)_n$ :
\begin{equation}
   (a\,;q)_n\eqdef\prod_{k=1}^n(1-aq^{k-1})=(1-a)(1-aq)\cdots(1-aq^{n-1}).
   \label{defqPoch}
\end{equation}
$\circ$ hypergeometric series ${}_rF_s$ :
\begin{equation}
   {}_rF_s\Bigl(\genfrac{}{}{0pt}{}{a_1,\,\cdots,a_r}{b_1,\,\cdots,b_s}
   \Bigm|z\Bigr)
   \eqdef\sum_{n=0}^{\infty}
   \frac{(a_1,\,\cdots,a_r)_n}{(b_1,\,\cdots,b_s)_n}\frac{z^n}{n!}\,,
   \label{defhypergeom}
\end{equation}
where $(a_1,\,\cdots,a_r)_n\eqdef\prod_{j=1}^r(a_j)_n
=(a_1)_n\cdots(a_r)_n$.\\
$\circ$ $q$-hypergeometric series (the basic hypergeometric series)
${}_r\phi_s$ :
\begin{equation}
   {}_r\phi_s\Bigl(
   \genfrac{}{}{0pt}{}{a_1,\,\cdots,a_r}{b_1,\,\cdots,b_s}
   \Bigm|q\,;z\Bigr)
   \eqdef\sum_{n=0}^{\infty}
   \frac{(a_1,\,\cdots,a_r\,;q)_n}{(b_1,\,\cdots,b_s\,;q)_n}
   (-1)^{(1+s-r)n}q^{(1+s-r)n(n-1)/2}\frac{z^n}{(q\,;q)_n}\,,
   \label{defqhypergeom}
\end{equation}
where $(a_1,\,\cdots,a_r\,;q)_n\eqdef\prod_{j=1}^r(a_j\,;q)_n
=(a_1\,;q)_n\cdots(a_r\,;q)_n$.\\
$\circ$ $q$-gamma function $\Gamma_q(z)$:
\begin{equation}
   \Gamma_q(z)\eqdef\frac{(q\,;q)_{\infty}}{(q^z\,;q)_{\infty}}
   (1-q)^{1-z},\qquad
   \lim_{q\nearrow 1}\Gamma_q(z)=\Gamma(z).
\end{equation}



\begin{thebibliography}{99}

\bibitem{askey}
G.\,E.\,Andrews, R.\,Askey and R.\,Roy,
{\it Special Functions\/},
Encyclopedia of mathematics and its applications, Cambridge, (1999).

\bibitem{ismail}
M.\,E.\,H.\,Ismail
{\it Classical and quantum orthogonal polynomials in one variable\/},
Encyclopedia of mathematics and its applications, Cambridge, (2005).

\bibitem{szego}
G.\,Szeg\"o,
{\it Orthogonal polynomials\/}, 4th edition,
Amer. Math. Soc. Colloq. Pub. {\bf 23} Providence, R.I. (1975).

\bibitem{infhul}
L.\,Infeld and T.\,E.\,Hull,
``The factorization method,''
Rev. Mod. Phys. {\bf 23} (1951) 21-68.

\bibitem{susyqm}
See, for example, a review:
F.\,Cooper, A.\,Khare and U.\,Sukhatme,
``Supersymmetry and quantum mechanics,''
Phys. Rep. {\bf 251} (1995) 267-385.

\bibitem{koeswart}
R.\,Koekoek and R.\,F.\,Swarttouw,
``The Askey-scheme of hypergeometric orthogonal polynomials and
its $q$-analogue,''
{\tt arXiv:math.CA/9602214}.

\bibitem{os12}
S.\,Odake and R.\,Sasaki,
``Orthogonal Polynomials from Hermitian Matrices,"
{\tt arXiv:\hspace{0pt}0712.4106[math.CA]}.

\bibitem{genden}
L.\,E.\,Gendenshtein,
``Derivation of exact spectra of the Schrodinger equation by means of
supersymmetry,''
JETP Lett. {\bf 38} (1983) 356-359.

\bibitem{crum}
M.\,M.\,Crum,
``Associated Sturm-Liouville systems,"
Quart. J. Math. Oxford Ser. (2) {\bf 6} (1955) 121-127,
{\tt arXiv:physics/9908019}.

\bibitem{os4}
S.\,Odake and R.\,Sasaki,
``Shape Invariant Potentials in `Discrete' Quantum Mechanics,''
J. Nonlinear Math. Phys. {\bf 12} Suppl. 1 (2005) 507-521,
{\tt arXiv:hep-th/0410102}.

\bibitem{os5}
S.\,Odake and R.\,Sasaki,
``Equilibrium Positions, Shape Invariance and Askey-Wilson Polynomials,''
J. Math. Phys. {\bf 46} (2005) 063513 (10 pages),
{\tt arXiv:hep-th/0410109}.

\bibitem{os6}
S.\,Odake and R.\,Sasaki,
``Calogero-Sutherland-Moser Systems, Ruijsenaars-Schneider-van Diejen
Systems and Orthogonal Polynomials,''
Prog. Theor. Phys. {\bf 114} (2005) 1245-1260,
{\tt arXiv:hep-th/0512155}.

\bibitem{os7}
S.\,Odake and R.\,Sasaki,
``Unified Theory of Annihilation-Creation Operators for Solvable
(`Discrete') Quantum Mechanics,''
J. Math. Phys. {\bf 47} (2006) 102102 (33pages),
{\tt arXiv:quant-ph/0605215};
``Exact solution in the Heisenberg picture and annihilation-creation
operators,"
Phys. Lett. {\bf B641} (2006) 112-117,
{\tt arXiv:quant-ph/0605221}.

\bibitem{os11}
S.\,Odake and R.\,Sasaki,
``$q$-oscillator from the $q$-Hermite Polynomial,''
{\tt arXiv:0710.\hspace{0pt}2209[hep-th]}.

\bibitem{rags}
O.\,Ragnisco and R.\,Sasaki,
``Quantum vs Classical Integrability in Ruijsenaars-Schneider Systems,''
J. Phys. {\bf A37} (2004) 469 - 479,
{\tt arXiv:hep-th/0305120}.

\bibitem{os3}
S.\,Odake and R.\,Sasaki,
``Equilibria of `discrete' integrable systems and deformations of
classical orthogonal polynomials,''
J. Phys. {\bf A37} (2004) 11841-11876,
{\tt arXiv:hep-th/\hspace{0pt}0407155}.

\bibitem{vD04}
J.\,F.\,van Diejen,
``On the Equilibrium Configuration of the $BC$-type Ruijsenaars-Schneider
System,"
J. Nonlinear Math. Phys. {\bf 12} Suppl. 1 (2005) 689-696,
{\tt arXiv:\hspace{0pt}math-ph/0410008}.

\bibitem{RS}
S.\,N.\,M.\,Ruijsenaars and H.\,Schneider,
``A new class of integrable systems and its relation to solitons,''
Annals Phys. {\bf 170} (1986) 370-405;
S.\,N.\,M.\,Ruijsenaars,
``Complete integrability of relativistic Calogero-Moser systems and
elliptic function identities,''
Comm. Math. Phys. {\bf 110} (1987) 191-213.

\bibitem{vD}
J.\,F.\,van Diejen,
``The relativistic Calogero model in an external field,''
{\tt arXiv:\hspace{0pt}solv-int/9509002};
``Multivariable continuous Hahn and Wilson polynomials related to
integrable difference systems,''
J. Phys. {\bf A28} (1995) L369-L374.

\bibitem{Stiel}
T.\,Stieltjes,
``Sur quelques th\'eor\`emes d'Alg\`ebre,"
Compt. Rend. {\bf 100} (1885) 439-440;
``Sur les polyn\^omes de Jacobi,"
Compt. Rend. {\bf 100} (1885) 620-622.

\bibitem{calnuovo}
F.\,Calogero,
``On the zeros of the classical polynomials,''
Lett. Nuovo Cim. {\bf 19} (1977) 505-507;
``Equilibrium configuration of one-dimensional many-body problems
with quadratic and inverse quadratic pair potentials,"
Lett. Nuovo Cim. {\bf 22} (1977) 251-253.

\bibitem{cs1}
E.\,Corrigan and R.\,Sasaki,
``Quantum vs Classical  Integrability in Calogero-Moser Systems,"
J. Phys. {\bf A35}  (2002) 7017-7061,
{\tt arXiv:hep-th/0204039}.

\bibitem{cal}
F.\,Calogero,
``Solution of the one-dimensional $N$-body problem with quadratic
and/or inversely quadratic pair potentials,''
J. Math. Phys. {\bf 12} (1971) 419-436.

\bibitem{sut}
B.\,Sutherland,
``Exact results for a quantum many-body problem in one-dimension. II,''
Phys. Rev. {\bf A5} (1972) 1372-1376.

\bibitem{st1}
R.\,Sasaki and K.\,Takasaki,
``Quantum Inozemtsev model, quasi-exact solvability and
${\cal N}$-fold supersymmetry,''
J. Phys. {\bf A34} (2001) 9533-9553,
Corrigendum J. Phys. {\bf A34} (2001) 10335,
{\tt arXiv:hep-th/0109008}.

\bibitem{os10}
R.\,Sasaki,
``Quasi Exactly Solvable Difference Equations,''
J. Math. Phys. {\bf 48} (2007) 122104 (11pages),
{\tt arXiv:0708.0702[nlin.SI]};
S.\,Odake and R.\,Sasaki,
``Multi-\hspace{0pt}Particle Quasi Exactly Solvable Difference Equations,"
J. Math. Phys. {\bf 48} (2007) 122105 (8pages),
{\tt arXiv:0708.0716[nlin.SI]}.

\bibitem{newqes}
R.\,Sasaki,
``New Quasi Exactly Solvable Difference Equation,"
{\tt arXiv:0712.2616\hspace{0pt}[nlin.SI]}.

\bibitem{Ush}
A.\,G.\,Ushveridze,
``Exact solutions of one- and multi-dimensional Schr\"{o}dinger
equations,''
Sov. Phys.-Lebedev Inst. Rep. {\bf 2}, 50, (1988) 54-58;
{\sl Quasi-exactly solvable models in quantum mechanics}
(IOP, Bristol, 1994);
A.\,Y.\,Morozov, A.\,M.\,Perelomov, A.\,A.\,Rosly, M.\,A.\,Shifman
and  A.\,V.\,Turbiner,
``Quasiexactly solvable quantal problems: one-dimensional analog of
rational conformal field theories,''
Int. J. Mod. Phys.  A {\bf 5} (1990) 803-832.

\bibitem{turb}
A.\,V.\,Turbiner,
``Quasi-Exactly-Solvable Problems and $sl(2)$ Algebra,"
Comm. Math. Phys. {\bf 118} (1988) 467-474.

\bibitem{vinzhed}
L.\, Vinet and A.\, Zhedanov,
``Quasi-linear algebras and integrability (the Heisenberg picture),"
SIGMA, to be published.

\bibitem{coherents}
T.\,Fukui and N.\,Aizawa,
``Shape-invariant potentials and an associated coherent state",
Phys. Lett. {\bf A180} (1993) 308--313;
J.-P.\,Gazeau and J.\,R.\,Klauder,
``Coherent states for systems with discrete and continuous spectrum",
J. Phys. {\bf A32} (1999) 123--132;
J.-P.\,Antoine, J.-P.\,Gazeau, P.\,Monceau, J.\,R.\,Klauder and
K.~A.\,Penson,
``Temporally stable coherent states for infinite well and
P\"oschl-Teller potentials",
J. Math. Phys. {\bf 42} (2001) 2349--2387;
A.\,H.\,El Kinani and M.\,Daoud,
``Coherent states \`a la Klauder-Perelomov for the P\"oschl-Teller
potentials",
Phys. Lett. {\bf A283} (2001) 291--299;
``Generalized coherent and intelligent states for exact solvable quantum
systems",
J. Math. Phys. {\bf 43} (2002) 714-733;
A.\,N.\,F.\,Aleixo and A.\,B.\,Balantekin,
``An Algebraic Construction of Generalized Coherent States for
Shape-Invariant Potentials",
J. Phys. {\bf A37} (2004) 8513, {\tt arXiv:quant-ph/0407160}.

\bibitem{degruij}
A.\,Degasperis and S.\,N.\,M.\,Ruijsenaars,
``Newton-Equivalent Hamiltonians for the Harmonic Oscillator,"
Ann. of Phys. {\bf 293} (2001) 92-109.

\end{thebibliography}
\end{document}